\documentclass[twocolumn,prb,hyperref,footinbib]{revtex4}
\usepackage{graphicx}
\usepackage{dcolumn}
\usepackage{bm}
\usepackage{bbm}
\usepackage{hyperref}
\usepackage{amsmath}
\usepackage{comment}
\usepackage{float}
\newcommand{\ket}[1]{\left|#1\right>}

\newcommand{\beq}{\begin{equation}}
\newcommand{\eeq}{\end{equation}}
\newcommand{\bea}{\begin{eqnarray}}
\newcommand{\eea}{\end{eqnarray}}
\newcommand{\nn}{\nonumber}
\newcommand{\tr}{\hbox{Tr}}

\begin{document}

\author{Sophia E. Economou$^1$ and Edwin Barnes$^{2}$}

\affiliation{$^{1}$Naval Research Laboratory, Washington, DC 20375, USA\\
$^{2}$Condensed Matter Theory Center and Joint Quantum Institute, Department of
Physics, University of Maryland, College Park, Maryland 20742-4111, USA}

\title{Theory of dynamic nuclear polarization and feedback in quantum dots}

\begin{abstract}
An electron confined in a quantum dot interacts with its local
nuclear spin  environment through the hyperfine contact interaction.
This interaction combined with external control and relaxation or
measurement of the electron spin allows for the generation of
dynamic nuclear polarization. The quantum nature of the nuclear
bath, along with the interplay of coherent external fields and
incoherent dynamics in these systems renders a wealth of intriguing
phenomena seen in recent experiments such as electron Zeeman
frequency focusing, hysteresis, and line dragging. We develop in
detail a fully quantum, self-consistent theory that can be applied
to such experiments and that moreover has predictive power. Our
theory uses the operator sum representation formalism in order to
incorporate the incoherent dynamics caused by the additional,
Markovian bath, which in self-assembled dots is the vacuum field
responsible for electron-hole optical recombination. The beauty of
this formalism is that it reduces the complexity of the problem by
encoding the joint dynamics of the external coherent and incoherent
driving in an effective dynamical map that only acts on the electron
spin subspace. This together with the separation of timescales in
the problem allows for a tractable and analytically solvable
formalism. The key role of entanglement between the electron spin
and the nuclear spins in the formation of dynamic nuclear
polarization naturally follows from our solution. We demonstrate the
theory in detail for an optical pulsed experiment and present an
in-depth discussion and physical explanation of our results.

\end{abstract}

\maketitle

\section{Introduction}

The electron-nuclear spin dynamics in quantum dots (QDs) have attracted
intense experimental and theoretical attention in recent years.\cite{Ono_PRL04,Bracker_PRL05,Koppens_Nature06,Greilich_Science07,Latta_NP09,Xu_Nature09,Vink_NP09,Kloeffel_PRL11,Merkulov_PRB02,Coish_PRB04,Deng_PRB06,Yao_PRB06,Cywinski_PRB09,Coish_PRB10,Erbe_PRL10,Faribault_PRL13} This is
both because of the role of the nuclear environment in potential
applications in quantum information, and because this is an
inherently interesting system that exhibits rich physics, especially
in the presence of external coherent and incoherent driving.

From a practical point of view, the nuclear spins comprise the main
source of electron  spin decoherence that limits the quality of spin
qubits. On the other hand, the ability to polarize the nuclear spins
allows them to be used as an asset instead of a liability. For
example, in the singlet-triplet qubit in electrostatically defined
quantum dots, the nuclear polarization is used as an effective
magnetic field to implement (psuedo)spin rotations.
\cite{Foletti_NP09} An ambitious role of the nuclear spins that
would take advantage of their long coherence times is their use as a
quantum memory, an idea that was proposed\cite{Taylor_PRL03} but not
yet demonstrated experimentally in quantum dots. Finally, an
additional motivation for gaining control over nuclear polarization
and controlling the nuclear spins is that a polarized and/or narrowed nuclear
bath polarization distribution would have less of a detrimental effect on the electron spin
coherence due to a reduction of fluctuations originating from a
reduced available phase space to which quantum information can be
lost.\cite{Rudner_PRL07,Barnes_PRL12b} In the case of gate-defined
QDs, it was demonstrated that significant amounts of nuclear
polarization or distribution narrowing can be generated and stabilized in a controlled
fashion,\cite{Baugh_PRL07,Bluhm_PRL10,Petersen_PRL13} and that this
can give rise to an enhancement of the spin coherence time by nearly
an order of magnitude.\cite{Bluhm_PRL10} In the context of
self-assembled QDs, a similar effect was achieved via coherent
population trapping, with an improvement in coherence time by a
factor of several hundred.\cite{Xu_Nature09}

From a fundamental science point of view, the \emph{open} and
\emph{driven}  electron-nuclear spin system is of great interest as
it has yielded a number of unexpected and intriguing
phenomena.\cite{Urbaszek_RMP13,Chekhovich_NM13} These arise from the
fact that driving the electron when it is coupled to a reservoir
(this can be, for example, a photon or phonon bath or cotunneling
with the leads) can produce dynamic nuclear spin polarization (DNP),
which in turn feeds back to the electron dynamics. This often
causes a reduction in nuclear spin fluctuations which manifests in a
variety of effects depending on the experimental setup. Noteworthy
phenomena include synchronizing of the electron spin frequency to
that of a periodic train of pulses, which can effectively homogenize
an ensemble of spins with a distribution of g-factors;
\cite{Greilich_Science07} locking of a driven optical transition to
the laser;\cite{Latta_NP09,Hogele_PRL12} hysteresis in the spectra
due to memory
effects.\cite{Xu_Nature09,Latta_NP09,Kloeffel_PRL11,Hogele_PRL12,Nilsson_PRB13}
There exist several theoretical works that analyze DNP processes  in
various experimental contexts. In the case of gate-defined QDs, a
range of phenomena have been studied such as DNP formation and
feedback,\cite{Deng_PRB05,Gullans_PRL10,Neder_arxiv13} nuclear spin
squeezing,\cite{Rudner_PRL11} dark state
formation,\cite{Gullans_PRL10,Gullans_PRB13} entanglement
dynamics,\cite{Rudner_PRL11,Schuetz_arXiv13} and dynamical
self-quenching.\cite{Brataas_PRL12} In the context of self-assembled
quantum dots, there exist several works that treat the problem of driving with a single continuous
laser that showed nuclear feedback effects and
hysteresis,\cite{Yang_PRB12,Yang_PRB13} as well as
for driving with two phase-locked pulses to achieve tunable polarization\cite{Shi_PRB13} and
nuclear spin cooling.\cite{Issler_PRL10}

\section{Overview of our approach}

Many of the experimental signatures of DNP repeat across
different setups in terms of driving sequences and charge configurations in the quantum dot. It is thus natural to seek a common
theoretical framework which can be adapted to explain any such type
of experiment.  In addition to the need for understanding existing
experimental results, a successful theory should also have
predictive power. The difficulty in setting up such a theory for
this system is the complexity of the problem: it is an \emph{open}
and \emph{driven} system which involves many degrees of freedom,
namely the electron spin, excited electronic states outside the
electron spin subspace,  the nuclear spins, and the reservoir that
causes the nonunitary dynamics. Moreover, there are feedback
effects: the generated nuclear spin polarization acts as an
effective magnetic field on the electron spin. Thus, the state of
the latter changes based on this updated magnetic field. The problem
clearly has to be solved self-consistently.

In this paper, we lay the foundations of such a theory, by expanding on
the formalism introduced in our earlier work. \cite{Barnes_PRL11} Our
theory is based on the use of dynamical maps. This is a powerful
tool that describes nonunitary evolution through operators that act
on the density matrix of the electron spin and evolve it in a
nonunitary fashion while preserving its trace. These operators are
found by solving for the dynamics of the electron system driven by
external fields and interacting with the reservoir. By solving for
the effect of these interactions on the electron spin, we can
eliminate any additional states outside the qubit subspace and the
degrees of freedom of the reservoir, while in principle accounting
for their effects  \emph{exactly}. This can allow for an analytical
approach that offers a  general, tractable and transparent treatment
of the problem.

Using the dynamical map that we find for the electron spin evolution
under the driving and coupling  to the reservoir, we calculate the
steady state electron spin vector, which constitutes our
zeroth-order solution (i.e., no coupling to nuclear spins). To
include nuclear effects, we perform a perturbative treatment on this
zeroth-order solution by finding the response of a single nuclear
spin to the motion of the electron spin under the external control.
We thus make the independent nuclear spin approximation. By
including the hyperfine coupling between the nucleus and electron,
we find the joint state of the two spins, which now includes quantum
correlations.

In this paper, we focus primarily on a large class of experiments
in which the driving is sufficiently fast that the electron spin reaches its
dynamical equilibrium steady state quickly compared to both the electron spin decoherence time and the
timescale of nuclear spin evolution. The so-called mode locking experiments\cite{Greilich_Science07,Carter_PRL09,Carter_PRB11}
are examples from this class, as will be demonstrated in the present work.\cite{foot1} For these types of experiments,
we can employ a Markovian approximation to separate the
nuclear spin degrees of freedom from those of the electron, which gives us an
effective dynamical map for the nuclear spin. The Markovian approximation is not only
valid when the electron dynamics are fast, it is also physically well motivated
by noting that when the relaxation to the steady state is fast compared to
decoherence, which is in turn fast relative to nuclear dynamics, the
electron spin will tend to remain in the steady state it attains in the absence of the nuclear spin. While the electron spin steady state
is approximately unaffected by a single nuclear spin, it will change significantly when the full nuclear spin ensemble is taken into account. This is explained in detail in the next paragraph.
Working in this Markovian limit, we obtain
an expression for the steady state of the nuclear spin which
explicitly involves all the parameters of the problem as well as the
electron steady state. To describe continuous wave driving and similar types of experiments, it may
be necessary to go beyond the Markovian limit.
However, the theory presented in this work can still be adapted to these cases as well, as
was done recently to explain experimental data for Ramsey fringes of hole spins, see Ref. \onlinecite{Carter_arxiv13}.

To take into account many-body multi-nuclear effects, we perform a
shift of the Zeeman frequency  of the electron by the total
effective magnetic field of all nuclear spins (Overhauser shift).
This is done by first finding a distribution for the nuclear spin
polarization using a mean field approach. We do this by solving a
kinetic equation that determines the probability $P(m)$ that the net
nuclear polarization is $m$. The quantity that enters in this
kinetic equation is the single-nucleus flip rate. Note that
generally the probability to flip from up to down is different than
that to flip from down to up. Both these rates are found by
solving the equation of motion of the single nuclear spin. With the
nuclear polarization distribution at hand, we then perform the
Overhauser shift and find the average steady state electron spin vector
self-consistently.

To explicitly demonstrate our formalism, in the second part of the
paper we focus on the spin mode locking
experiments\cite{Greilich_Science07,Carter_PRL09} in which a train of
fast circularly polarized pulses is applied to the electron.  We
show that our theory reproduces the main experimental features,
including the buildup of nuclear spin polarization and its role in
electron spin frequency synchronization\cite{Greilich_Science07} and
antisynchronization\cite{Carter_PRL09} with the pulse train. Furthermore,
in this work we go beyond the high magnetic field approximation of
Ref. \onlinecite{Barnes_PRL11} by taking into account the so-called
spontaneously generated coherence phenomenon,\cite{Javanainen_EL92} which
strongly modifies the generation of electron spin polarization at
low magnetic fields.\cite{Economou_PRB05,Dutt_PRL05} We find that in this
regime there is larger nuclear spin polarization compared to the
higher magnetic field case, but that it takes a longer time to reach
the steady state.

We also examine modifications to the mode locking experimental
setup. In particular, we calculate the dynamics when an additional,
coherent spin echo pulse is included in each period. Such pulses are
important in the context of quantum information as they constitute
the simplest form of dynamical decoupling. We show that this pulse
sequence leads to strong electron spin polarization in the plane
transverse to the magnetic field, modifies the synchronization
effect, and overall reduces the average nuclear spin polarization.

This paper is organized as follows. In Section \ref{sec:dnpQDs} we
give an intuitive explanation of DNP in terms of electron-nuclear
spin entanglement. In Section \ref{sec:kraus} we present a brief
review of the operator sum formalism, and in Section
\ref{sec:generalformalism} we motivate and review our general
formalism. Section \ref{ML} is devoted to analyzing and explaining
the pulsed mode locking experiments.\cite{Greilich_Science07, Carter_PRL09}

\section{Dynamic nuclear polarization in quantum dots} \label{sec:dnpQDs}

Dynamic nuclear polarization is nuclear polarization generated
through dynamic processes, most commonly external driving fields and
some kind of incoherent dynamics of the electron spin,  instead of
by simple nuclear spin relaxation (cooling) to a polarized ground
state. Overhauser was the first to predict such an effect in the
early 1950s,\cite{Overhauser_PR53}, and his prediction was originally
met with skepticism until it was verified by Slichter and
Carver.\cite{Carver_PR56} Since then, there has been a huge
number of DNP experiments conducted in a variety of diverse systems
and based on various nonunitary physical processes. A key component
of DNP is clearly a mechanism that removes entropy from the system.
Such nonunitary mechanisms may correspond to relaxation or
measurement. In the case of self-assembled quantum dots the
experiments are optical and involve an excited state outside the
electron spin subspace, typically a charged exciton, created by a
(quasi) resonant laser focused on the band gap. The extraction of
entropy from the system happens through optical excitation followed
by recombination and spontaneous emission of a photon. The emitted
photon generally carries information about the system and can
therefore lower the entropy of the net electron-nuclear state.
Experimentally, this can coincide with the actual measurement of the
system, but this is not necessarily always the case.

Optical experiments in QDs have revealed a distinct incarnation of
the DNP effect  and rich physics based on the interplay of the
optical driving, the spontaneous recombination and of course the quantum
many-body nuclear bath. It is perhaps useful at this
point to discuss what distinguishes these DNP experiments with QDs
from more conventional DNP demonstrations. One feature of the QD is
that it involves a large number of nuclear spins, about $10^4-10^6$,
depending on dot size. Therefore the nuclear spectra can be thought
of as bands instead of discrete energy levels and generally cannot
be resolved by the external fields.

A more important feature however is the role of nuclear feedback. As
mentioned above,  nuclear polarization acts as an effective magnetic
field that shifts the Zeeman frequency of the electron spin. The
distinctive feature in QD experiments is that there exist
selection rules which affect electron spins differently depending on their
orientation and energy. Therefore a shift in the Zeeman splitting is
not just a small quantitative correction, but can instead change
\emph{qualitatively} the behavior of the system. For example, in the
optical mode locking experiment an electron with a Larmor period that
is an integer multiple of the pulse repetition period will become
fully polarized and will subsequently be insensitive to the pulse
due to polarization selection rules. On the other hand, an electron
with a Larmor period that is a half integer multiple of the period
will be minimally polarized by the pulse train. This example
demonstrates how the nuclear feedback can have a large effect on the
behavior of the electron spin and why a self-consistent treatment is
necessary.

To close this section let us present the physical picture of DNP
generation in QDs via a toy model.\cite{foot2}  Consider two spins, one
initialized in a pure state and the other in a mixed state,
corresponding to the electron and nuclear spin respectively. Now
allow them to evolve under a Heisenberg type interaction $A
\mathbf{S}_1 \cdot \mathbf{S}_2$. The evolution from the initial
state to the state at time $t=\pi/A$ is described as
\begin{eqnarray}
|\uparrow\rangle \langle \uparrow| {\otimes}(|\uparrow\rangle \langle
\uparrow|{+}|\downarrow\rangle \langle \downarrow|)  \rightarrow
(|\uparrow\rangle \langle \uparrow|{+}|\downarrow\rangle \langle
\downarrow|){\otimes} |\uparrow\rangle \langle \uparrow|. \nonumber
\end{eqnarray}
In the language of quantum information, we can view this as a swap
gate,  meaning that the two spins have swapped
quantum states. Now at $t=\pi/A$ a pulse comes in which performs a
projective measurement on the first spin and collapses it, e.g.,
into state $|\downarrow\rangle$. This process leaves \emph{both}
spins in a pure (i.e., fully polarized) state even though the
nuclear spin never interacted directly with the external field. This
is precisely the process that removes entropy from the system via
the measurement. The purpose of this toy model is to demonstrate
this effect in a straightforward manner and hopefully build
intuition into the more complicated dynamics that we present below.

\section{Operator sum representation (Kraus) formalism} \label{sec:kraus}

In quantum mechanics, a closed system undergoes unitary evolution.
However, that is not the most general type of evolution. A system
generally interacts with other systems, and energy and entropy can
be exchanged with  them through this interaction. The system is then
called open, and the operator sum representation formalism can be
used to describe its nonunitary evolution. The
operators that describe this  irreversible evolution are called
Kraus operators,\cite{Nielsen_Chuang} and they act on a density matrix $\rho$ in the
following way
\begin{eqnarray}
\rho^\prime = \sum_k E_k \rho E^\dag _k,
\label{rhokrausgeneric}
\end{eqnarray}
where $k>1$ and the relation $$\sum_k E^\dag _k E_k   = \mathbbm{1}$$
(where $\mathbbm{1}$ is the identity operator)
should hold in order to  guarantee that the trace of the density
matrix remains equal to one. As a simple example, consider a
two-level system where the population can relax from the excited to
the ground state irreversibly. This is an ubiquitous scenario across
physical systems, e.g., this may be an atom in a metastable
optically excited state, or a nuclear spin, etc. The Kraus operators
that describe the decay from the excited to the ground state are (in
the basis \{$|g\rangle , |e\rangle$\})
\begin{eqnarray}
M_0  &=& \left[
\begin{array}{cc}
1 & 0 \\
0 & \sqrt{\alpha}
\end{array}
\right], \nonumber
\\
M_1  &=& \left[
\begin{array}{cc}
0 & \sqrt{1-\alpha} \\
0 & 0
\end{array}
\right].
\end{eqnarray}
Starting from an arbitrary initial density matrix,
\begin{eqnarray}
\rho &=& \left[
\begin{array}{cc}
\rho_{11} & \rho_{12} \\
\rho_{21} & \rho_{22}
\end{array}
\right],
\end{eqnarray}
the final density matrix after the probabilistic decay process has completed is then
\begin{eqnarray}
\rho^\prime &=& \left[
\begin{array}{cc}
\rho_{11} + (1-\alpha)\rho_{22} & \sqrt{\alpha}\rho_{12} \\
\sqrt{\alpha}\rho_{21} & \alpha\rho_{22}
\end{array}
\right].
\end{eqnarray}
It is simple to check that when $\alpha = 0$ we have complete
relaxation from the excited  to the ground state, while $\alpha=1$
yields the trivial solution of no decay, with the system remaining
in its initial state without evolving. It is useful to note here
that one could make $\alpha$ a time dependent parameter. In that
case, the density matrix can be found at any time using
Eq.~(\ref{rhokrausgeneric}). For exponential decay, we would have
$\alpha= \alpha (t)= e^{-t/T_1}$, while the decoherence, described by the decay of the off-diagonal density matrix component, occurs with a
timescale $T_2=2 T_1$ as it should.

\section{General formalism}\label{sec:generalformalism}

The total Hamiltonian of the system is
\begin{eqnarray}
H(t)=H_{0,e}+H_c(t)+H_{res}+H_{0,n}+H_{hf}, \label{genham}
\end{eqnarray}
where $H_{0,e}$ is the free part of the electron Hamiltonian in the
QD, $H_c(t)$ is the control Hamiltonian, and $H_{res}$ is the
interaction with the reservoir.  We will focus on the case where
$H_c(t)=H_c(t+T_R)$ describes a periodic sequence of finite-duration pulses with a period $T_R$. In
general, these pulses will couple the electron spin states to higher
excited levels. All the population decays back to the electron spin
subspace through the interaction with the reservoir, $H_{res}$, with characteristic rate $\gamma$. In
self-assembled dots, $H_c(t)$ describes optical pulses coupling the
electron spin states to additional levels that are charged excitons,
also called trions, and $H_{res}$ is the photon bath. In
electrostatically defined QDs, $H_c(t)$ is a gate voltage, the
additional states can be, for example, the two polarized triplet
states that lie outside the singlet-triplet qubit subspace, and
$H_{res}$ represents the interaction with the leads. The remaining two
terms are the nuclear spin Hamiltonian in the presence of a magnetic
field, $H_{0,n} =\omega_n {\sum}_i \hat{I}^i_z$, and the hyperfine
interaction between the electron and $N$ nuclei,
\begin{eqnarray}
H_{hf} = \sum _{i=1}^N A_i \hat{S}_z \hat{I}^i_z + \sum _{i=1}^N
{A_i}/{2}(\hat{S}_+\hat{I}^i_-+\hat{S}_-\hat{I}^i_+).\label{hfham}
\end{eqnarray}
The first term in $H_{hf}$ is referred to as the Overhauser term, while
the second is known as the flip-flop term. The hyperfine couplings are
determined by the magnitude of the electronic wavefunction at the locations
of the nuclear spins: $A_i={\cal A}\text{v}_0|\Psi(r_i)|^2$, where $\cal{A}$ is
the total hyperfine energy, $\text{v}_0$ is the volume per nucleus, $\Psi$ is the
electronic wavefunction, and $r_i$ is the location of the $i$th nucleus.

There are two features of the open electron-nuclear spin system that are
advantageous toward the development of a formalism to treat this problem.
The first feature is that the control Hamiltonian, $H_c(t)$, acts solely
on the electron spin subsystem and does not directly affect the nuclear
spins. This fact, combined with the smallness of the hyperfine couplings
compared with the electron Zeeman frequency, allows us to first solve for
the electron evolution in the absence of the nuclei and to then compute
the response of the nuclear spins to the electron dynamics. Specifically, we
employ a perturbative expansion in the hyperfine flip-flop interaction to
obtain analytical expressions for the nuclear steady state and relaxation
rate.

The second useful feature is a hierarchy of timescales. In particular,
we primarily focus on experiments in which the reservoir-induced relaxation from
auxiliary excited states to the electron spin subspace is fast compared
to the driving period: $\gamma T_R\gg1$. This will allow us to describe
the evolution over one period in terms of a dynamical map that acts only
on the 2$\times$2 electron spin subspace instead of a larger
dimensional Hilbert space. This in turn enables us to
coarse-grain the electron spin evolution by piecing together copies of
this dynamical map, leading to a substantial simplification of the
analysis, and allowing for greater insight into the physics. We
also take advantage of a second timescale hierarchy, namely
$\tau_e\ll T_2\ll \tau_n$, where $\tau_e$ is the time it takes for the
electron to reach its steady state, $T_2$ is the decoherence time of the
electron spin, and $\tau_n$ is a characteristic timescale for nuclear dynamics.
In this regime, decoherence works to keep the electron spin in the steady state
it would have in the absence of nuclei, although the nuclear Overhauser field
will still induce a shift in the electron Zeeman frequency. This indicates that a
Markovian approximation in which electron-nuclear correlations are discarded
after each driving period is not only justified but physically
well motivated. Note that, even in systems where these timescale
hierarchies do not hold (such as in singlet-triplet qubits), so long as there is
a nonunitary process that resets the qubit, we would still expect a Markovian
approximation to apply. One difference, however, is that instead of first
calculating the electron spin steady state alone, one may need to calculate the
total electron-nuclear spin (nonunitary) evolution per cycle.

In the following subsections, we describe in detail our general formalism
as it applies to the large class of experiments exhibiting the timescale
hierarchies described above. The first step is to derive the dynamical map
describing the evolution of the electron system without the hyperfine
interactions, and to use this result to compute the electron spin steady
state. We then couple a single nuclear spin to the electron and calculate
its resulting steady state and relaxation rate. These quantities are the
ingredients needed to construct the multi-nuclear flip rates that enter into
a kinetic equation for the nuclear spin polarization distribution of the entire
nuclear spin ensemble. The solution of this kinetic equation then gives the
polarization distribution generated by a particular driving sequence.
Finally, we obtain the nuclear feedback on the electron spin
steady state by performing an Overhauser shift in the Zeeman frequency and averaging
the resulting modified steady state over the polarization distribution. In the second half of the paper, we apply
our formalism to the particular case of the mode locking experiments.\cite{Greilich_Science07, Carter_PRL09}
We demonstrate explicitly the requisite hierarchy of timescales, and we show that our
formalism reproduces the salient features of the experimental findings.

\subsection{Electron spin Kraus operators} \label{sec:espinKraus_gen}

To find the zeroth-order solution as a 2$\times$2 operation on the electron
spin only, we first take the standard approach of treating the
reservoir to second order under  the Markovian approximation, which
gives rise to decay and decoherence terms in the Liouville-von Neumann
equation. These terms can be described by Lindblad operators so
that, ignoring nuclear terms and defining $H_e(t)=H_{0,e}+H_c(t)$, the
total evolution for the electron subsystem is described by
\begin{eqnarray}
\dot R=i[R,H_{e}(t)] + \mathcal{L}(R),
\label{LvNeqn}
\end{eqnarray}
where the symbol $R$ is used to stress that this density matrix
includes  the two spin states \emph{and} the excited states that
couple to the spin subspace via $H_c(t)$. It is important to note that the initial condition for (\ref{LvNeqn}) is an arbitrary density matrix \emph{in the spin subspace}, i.e., only a 2$\times$2 block of nonzero
matrix elements. Since we are interested only in the spin subspace,
we would like to use Eq. (\ref{LvNeqn}) to construct a dynamical map that
describes only the evolution of this subspace in terms of 2$\times$2 matrices.
To facilitate this construction, we focus on the regime in which the relaxation is fast compared
to the pulse period ($\gamma T_R\gg 1$). Since the theory can be applied for multiple pulses per period, a more precise condition would in fact be that $1/\gamma$ should be small compared to the largest time delay between pulses that occurs in the pulse sequence. In this case, the density matrix $R$ after one period is such that the components outside the
2$\times$2 spin subspace block are negligibly small, and we can derive a dynamical map that evolves the spin subspace over one period $T_R$.
 For one or even two excited states, this
can often be done analytically. Otherwise, a perturbative or
numerical approach is needed. The solution either way will provide
an expression for $\rho^\prime$, the density matrix of the electron
spin after one period, as a function of the
initial density matrix $\rho$. From this we can extract the Kraus
operators $\{\mathcal{E}_k\}$  since they are used to relate $\rho^\prime$ to
$\rho$:
\begin{eqnarray}
\rho^\prime =\sum_k \mathcal{E}_k \rho \mathcal{E}_k^\dagger.
\label{transformation}
\end{eqnarray}
Note that the $\{\mathcal{E}_k\}$ contain the evolution of the whole period, including both the unitary part due to the free Hamiltonian and the coherent control effects and the nonunitary part due to the reservoir. The explicit form of the Kraus operators
for pulsed experiments will be given in Section \ref{ML}.

\subsection{Spin vector representation} \label{sec:SVrep_gen}

For the present problem, the density matrix is not a convenient
representation of the spin state. The reason is that to find the
steady state of the electron spin, we need to operate on it with the
appropriate Kraus operators an infinite number of times, and since
these operators act on both sides of the density matrix, this
quickly becomes intractable. A much more convenient way to solve
this problem is to transform to the spin vector (SV) representation,
which is a completely equivalent way of representing the state of
the system, but with the important property that the operators
describing the evolution act on the left only. \cite{Nielsen_Chuang}
In addition, the spin vector representation offers a compelling
geometric visualization of the dynamics.

Before we proceed with the derivation of the SV
representation from Eq.~(\ref{transformation}), let  us first
discuss what kind of physics the dynamical map of the spin should
describe. Obviously, the evolution will generally be nonunitary, but
what does that mean for an input state? Clearly, a pure state
undergoing nonunitary evolution will generally \emph{lose} purity
and will become (partially or fully) mixed. Note however that a
mixed state may either become \emph{more} or \emph{less} mixed under
nonunitary evolution. The latter case, where the system gains
purity, is equivalent to increasing the spin polarization in the system.
In the special case of zero initial polarization, the pulse and
subsequent reservoir-induced relaxation will generate a nonzero spin
vector after one driving period. We thus expect the general form of the evolution of the spin
vector $S$ over one period to be given by
\begin{eqnarray}
S^\prime = Y S + K,
\label{svprime}
\end{eqnarray}
where $S$ and $S^\prime$ correspond to density matrices $\rho$ and
$\rho^\prime$ respectively  in Eq. (\ref{transformation}). We define the spin vector to be normalized to unity, i.e., its components are given by $S_m=\tr(\rho\sigma_m)$, where $\sigma_m$ denotes the Pauli matrices. In general, the matrix $Y$ both rotates and shrinks
the spin vector due to population loss, while $K$ restores the population
to the electron spin Hilbert space. To find
$Y$ and $K$ we start from the general equation
\begin{eqnarray}
\rho^\prime = \sum_j \mathcal{E}_j \rho \mathcal{E}_j^\dag
\end{eqnarray}
and multiply both sides by the Pauli matrix $\sigma_\ell$ and take
the  trace:
\begin{eqnarray}
\text{Tr}(\sigma_\ell\rho^\prime) = \text{Tr}\left(\sum_j \sigma_\ell \mathcal{E}_j \rho \mathcal{E}_j^\dag \right).
\end{eqnarray}
The LHS is just $S_\ell$, and we express $\rho$ on the RHS in terms
of the SV, i.e., we make  the substitution $\rho=1/2+1/2\sum_m
\sigma_m S_m$ to obtain:
\begin{eqnarray}
S_\ell^\prime =  K_\ell + \sum_m Y_{\ell,m}  S_m,
\end{eqnarray}
where
\begin{eqnarray}
K_\ell &=& \frac{1}{2} \text{Tr} \sum_j \sigma_\ell \mathcal{E}_j  \mathcal{E}_j^\dag = \text{Tr} \sum_j s_\ell \mathcal{E}_j  \mathcal{E}_j^\dag, \label{K3}\\
Y_{\ell,m} &=& \frac{1}{2}\text{Tr} \sum_j \sigma_\ell \mathcal{E}_j \sigma_m \mathcal{E}_j^\dag = 2 \text{Tr} \sum_j s_\ell \mathcal{E}_j s_m
\mathcal{E}_j^\dag,
\label{Y3}
\end{eqnarray}
where we define $s_j=\frac{1}{2} \sigma_j$.

\subsection{Zeroth-order solution: the steady state electron spin vector} \label{sec:zerothordersln_gen}

An unpolarized spin undergoing the evolution described by $Y$ and $K$ will
obtain some  polarization. The spin right after the first, second,
and $n$th driving period will be
respectively
\begin{eqnarray}
S_{1} &=&  K \nonumber \\
S_2 &=& Y S_1 + K  =  Y K + K  \nonumber \\
S_n &=&  Y S_{n-1} + K  =  \left( Y^{n-1}+...+ Y + \mathbbm{1} \right) K . \label{geomseries}
\end{eqnarray}
Eq. (\ref{geomseries}) is a geometric series; we can
therefore readily write down the expression for the steady  state
spin vector \emph{at the end} of a driving period as
\begin{eqnarray}
S_\infty = (\mathbbm{1}-Y)^{-1} K.
\end{eqnarray}
The inverse in the above equation in general exists because the
eigenvalues of $Y$ are all  less than unity, as follows from the fact that $Y$ includes the
loss of population to the excited state. Before we proceed to the
inclusion of the nuclear spin,  we will consider a slightly
modified, but equivalent, version of this formalism, where the
vector $K$ and the matrix $Y$ are represented by a single 4$\times$4 matrix:
\begin{eqnarray}
\mathcal{Y}_e=\left[
\begin{array}{cccc}
1 & 0 & 0 & 0 \\
K_{x} & Y_{xx} & Y_{xy} & Y_{xz} \\
K_{y} & Y_{yx} & Y_{yy} & Y_{yz} \\
K_{z} & Y_{zx} & Y_{zy} & Y_{zz}
\end{array}
\right]. \label{espin}
\end{eqnarray}
It is easy to check that in this 4d representation, the steady-state SV
$\mathcal{S}^{(\infty)}_{e}=(1,S^{(\infty)}_{e,x},S^{(\infty)}_{e,y},S^{(\infty)}_{e,z})
$ is the eigenvector of $\mathbbm{1}-\mathcal{Y}_e$ with eigenvalue zero. It is generally
the case that the first component of the 4d spin vector must remain fixed at 1 in order
for $\mathcal{Y}_e$ to evolve the remaining three components of the spin vector appropriately. This
more compact representation will prove very useful when we
introduce the nuclear spin.

\subsection{Including a single nuclear spin} \label{sec:includesinglesnspin_gen}

The next goal is to find an equation similar to Eq. (\ref{svprime}) for the nuclear spin, and
from that derive the steady state nuclear spin vector along with the relaxation rate. These quantities
will later be  used as inputs into the equation that determines the nuclear polarization distribution for the
entire ensemble of $N$ nuclear spins. For
simplicity, we focus on the case of spin 1/2 nuclei throughout the paper, but the
formalism could be extended to consider other species of nuclei as well.

We begin by finding the appropriate Kraus operators for the two-spin
system. Here we are keeping them arbitrary since  we are interested
in presenting the general method, but later in Section \ref{ML} we
will derive the Kraus explicitly for the pulsed problem. Defining
the two-spin Kraus operators as $\mathcal{F}_j$, we evolve the density matrix $\mathcal{P}$ describing the total electron-nuclear spin state over one driving period according to
\begin{eqnarray}
\mathcal{P}^\prime = \sum_j \mathcal{F}_j \mathcal{P}\mathcal{F}_j^\dag.
\end{eqnarray}

Let us now define generalized Pauli matrices for the two-spin system, which are tensor products
of the usual Pauli matrices, including unity:
\begin{eqnarray}
G_{4k+\ell}=s_k\otimes s_\ell,
\end{eqnarray}
where $k,\ell$ run from 0 to 3, with $s_0 \equiv \frac{1}{2}\mathbbm{1}$. Using these
operators, we can define the spin vector for the joint system. There
are 16 different $G$'s, but only 15 numbers are needed to specify
the state due to the normalization constraint. However, in analogy
to the 4d SV representation defined above, we work in a 16d representation in which $\mathcal{S}$
denotes the two-spin SV containing both the electronic and nuclear spin degrees of freedom, i.e., $\mathcal{S}_i=4\tr(\mathcal{P} G_i)$.
In general, $\mathcal S$ is not simply a tensor product of the two individual SVs, but contains quantum
correlations between the electron and nuclear spins. In this representation,
the evolution operator over one period is given by
\begin{eqnarray}
\mathcal{Y}_{ij}=4\sum_\ell\text{Tr}\left[G_i \mathcal{F}_\ell  G_j \mathcal{F}_\ell^\dagger\right],
\end{eqnarray}
with the total spin vector evolving according to
\begin{eqnarray}
\mathcal{S}' = \mathcal{Y} \mathcal{S}.
\end{eqnarray}

In principle, we could obtain the two-spin steady state by finding the eigenvector of
$\mathbbm{1}-\mathcal{Y}$ with vanishing eigenvalue in direct analogy with the single electron spin case
treated above. However, we will instead perform a Markovian approximation which amounts  to keeping only the
separable (tensor product) part of $\mathcal{S}$, i.e.,
$\mathcal{S}\approx \mathcal{S}^{(\infty)}_{e}\otimes\mathcal{S}_n$.
As discussed above, this approximation is valid when there is a separation of timescales, in particular when the electron reaches its steady state,
$\mathcal{S}^{(\infty)}_{e}$, quickly compared to the
nuclear dynamics and the electron spin decoherence time. When this is the case, the electron tends to remain in the steady state it would have without interactions with the nuclei, suggesting that the Markovian treatment is in fact more physical.  We then obtain
the effective nuclear spin evolution $\mathcal{Y}_n$ by acting with
$\mathcal{Y}$ on the tensor product
$\mathcal{S}^{(\infty)}_{e}\otimes\mathcal{S}_n$ and reading off the
coefficients of the components of $\mathcal{S}_n$ from the resulting $\mathcal{S}'$. This procedure
can be summarized by the equation
\begin{eqnarray}
(\mathcal{Y}_n)_{\alpha\beta}=\frac{d}{d\mathcal{S}_{n,\beta}}\left[ \mathcal{Y} (\mathcal{S}^{(\infty)}_{e}\otimes\mathcal{S}_n) \right]_\alpha,
\end{eqnarray}
where the resulting $\mathcal{Y}_n$ explicitly contains electron SV components. From $\mathcal{Y}_n$ we
find the nuclear spin steady  state
$\mathcal{S}^{(\infty)}_{n}=(1,S^{(\infty)}_{n,x},S^{(\infty)}_{n,y},S^{(\infty)}_{n,z})$
as the eigenvector of $\mathbbm{1}-\mathcal{Y}_n$ with eigenvalue equal to zero.

Next, we explain how to derive the nuclear relaxation rate. The evolution of the 4d nuclear SV is
described by
\begin{eqnarray}
\mathcal{S}_n(t+T_R) = \mathcal{Y}_n\mathcal{S}_n(t).
\end{eqnarray}
Since the nuclear evolution is much slower
than $T_R$, we can coarse grain this equation to
obtain a differential equation for the nuclear SV:
\begin{eqnarray}
\frac{d}{dt} \mathcal{S}_n = {1\over T_R}(\mathcal{Y}_n-\mathbbm{1})\mathcal{S}_n,
\end{eqnarray}
which gives
\begin{eqnarray}
\mathcal{S}_n(t)=e^{(\mathcal{Y}_n-\mathbbm{1}) t/T_R}\mathcal{S}_n(0).\label{coarseSn}
\end{eqnarray}
It is clear from this result that the smallest nonzero eigenvalue, $\lambda_2$, of $\mathbbm{1}-\mathcal{Y}_n$
will determine the relaxation rate of the nuclear spin: $\gamma_n=\lambda_2/T_R$.

\subsection{Nuclear spin steady state and relaxation rate in the perturbative regime} \label{sec:nucspinSS_gen}

In the previous subsection, we showed that in the Markovian limit, the nuclear steady state and relaxation rate can be obtained from the effective evolution operator (in the SV representation) for a single nuclear spin over one period, $\mathcal{Y}_n$. Specifically, the steady state is given by the eigenvector of $\mathbbm{1}-\mathcal{Y}_n$ with eigenvalue zero, while the relaxation rate is inversely proportional to the smallest nonzero eigenvalue of $\mathbbm{1}-\mathcal{Y}_n$. In order to obtain explicit analytical results, we make use of the fact that the hyperfine couplings are small compared to the electron Zeeman energy and perform a perturbative expansion in the hyperfine flip-flop interaction. We keep the Overhauser part of the interaction to all orders in the coupling (see Eq. (\ref{hfham})). In Appendix~\ref{app:nucspinSS_gen}, we show that to leading order in this perturbative expansion, the nuclear spin steady state has the form
\beq
\mathcal{S}^{(0)}_{n}=(1,0,0,\xi^*),\label{nucss}
\eeq
where the nuclear spin components transverse to the magnetic field vanish to leading order. It is further shown in the appendix how to explicitly calculate $\xi^*$ as well as the smallest nonzero eigenvalue, $\lambda_2^*$, of $\mathbbm{1}-\mathcal{Y}_n$. These quantities will be used below to determine the nuclear spin flip rates.

\subsection{Driving with a simple periodic pulse train} \label{sec:drivingwsimpleperiodicpulse_gen}

An important class of driving sequences involves a periodic pulse train
with a single pulse per period. This includes, but is not limited
to, the case of mode locking,  which will be analyzed in depth
below. For this class of driving sequences, the explicit expressions
for each term in the perturbative hyperfine expansion of $\mathcal{Y}_n$ up to second order
are given in Appendix~\ref{app:Yn}, and the full expressions for
$\xi^*=S^{(\infty)}_{n,z}$ and $\lambda_2^*$ are given in
Appendix~\ref{app:nssrr}. In the remainder of the paper, we will
denote the nuclear spin steady state by $S^{(\infty)}_{n}$. In the limit $\omega_n\to0$, the
expressions for $S^{(\infty)}_{n,z}$ and $\lambda_2^*$ reduce to the
results quoted in Ref.~[\onlinecite{Barnes_PRL11}]: \beq
S^{(\infty)}_{n,z}=\frac{S_{e,z} \left[\left(S_e^2-1\right) \cos
\left(\frac{A
T_R}{2}\right)+S_e^2+1\right]}{S_{e,z}^2+\left(S_e^2-1\right) \cos
\left(\frac{A
   T_R}{2}\right)+1},\label{snzwn0}
\eeq \beq \lambda_2^*=\frac{A^2}{\omega_e^2}
\frac{1+S_{e,z}^2+(S_e^2-1)\cos(\frac{A
T_R}{2})}{1+S_{e,z}^2+(S_{e,z}^2-1)\cos(\frac{A
T_R}{2})}\sin^2\frac{\omega_e T_R}{2}.\label{lambda2wn0} \eeq In the
above expressions, we have compressed the notation for the electron
steady state $S_{e,i}^{(\infty)}\to S_{e,i}$ for  the sake of
brevity, and we have defined $S_e^2\equiv
S_{e,x}^2+S_{e,y}^2+S_{e,z}^2$.

Given the generality of Eqs.~(\ref{snzwn0})-(\ref{lambda2wn0}) it is
worth pausing for a moment to examine the physical content  of these
expressions. First, the fact that $S_{n,z}^{(\infty)}$ is
proportional to $S_{e,z}^{(\infty)}$ is a reflection of conservation
of angular momentum, which requires that $S_{n,z}^{(\infty)}=0$ when
$S_{e,z}^{(\infty)}=0$. Second, it can be seen from
Eq.~(\ref{lambda2wn0}) that when $AT_R\ll1$ (as is typically
necessary for the validity of the Markovian approximation) and when
the electron spin is polarized primarily along the directions
transverse to the magnetic field (e.g. $S_{e,x}^{(\infty)}\approx
1$), $\lambda_2^*$ and hence $\gamma_n$ become very large, leading
to rapid flipping of the nuclear spin. This behavior can be
attributed to the fact that the electron spin flips more easily when
it is polarized transversely to the B-field since in this case
hyperfine flip-flops do not violate energy conservation. On the
other hand, when the electron spin is polarized along the magnetic
field direction, flip-flops are suppressed due to the large Zeeman
energy mismatch between the electron and nuclear spins.

Third, we point out the factor $\sin^2\frac{\omega_e T_R}{2}$ in
Eq.~(\ref{lambda2wn0}), which indicates the importance of the
driving period relative to the Zeeman frequency. In particular, we
would like to address why the rate is zero when the electron spin
has a precession period that is commensurate with the driving
period, while it is maximized when the precession period is a half
integer multiple of the driving period. To understand this, we
consider a simple model in which the driving is a train of pulses,
each acting on one of the two electron spin states along the
magnetic field, i.e., the eigenstates of the  free electron
Hamiltonian, and exciting that state to an auxiliary, trion level.
For concreteness we choose to drive the spin down state,
$\ket{\downarrow}$. We consider the simplest case of an
instantaneous, resonant pulse, such that the electron spin steady state in
the absence of nuclei is simply the other spin state,
$\ket{\uparrow}$, i.e., $S_{e,z}=S_{e}=1$. Plugging these values
into Eq.~(\ref{lambda2wn0}) we obtain the simple expression
$\lambda_2^*=\frac{A^2}{\omega_e^2} \sin^2\frac{\omega_e T_R}{2}$.
Now we consider adding a nuclear spin in order to see physically the
origin of this expression. In particular, under the Heisenberg-type
interaction, the evolution operator of the two-spin system (electron
and nuclear spin) after one period is $U_{hf}(T_R)=e^{-i (H_{0,e}+H_{hf})T_R}$.

We consider the two limiting cases mentioned above,
$T_R=2n\pi/\omega_e$ and $T_R=(2n+1)\pi/\omega_e$. Expanding  the
corresponding evolution operators to second order in $A/\omega_e$,
and applying them to a state with $S_{e,z}=S_{e}=1$ and an arbitrary
nuclear spin state, i.e., $\ket{\uparrow}(c_\uparrow \ket{\uparrow}
+ c_\downarrow \ket{\downarrow})$ we obtain: When
$T_R=\frac{2n\pi}{\omega_e}$
\begin{eqnarray}
\ket{\uparrow}(c_\uparrow e^{-i\pi \frac{A}{\omega_e}} \ket{\uparrow} + c_\downarrow e^{i\pi \frac{A}{\omega_e}} \ket{\downarrow}),\label{upupevol2pi}
\end{eqnarray}
and when $T_R=\frac{(2n+1)\pi}{\omega_e}$
\begin{eqnarray}
c_\uparrow e^{-i\pi\frac{A}{2\omega_e}}  \ket{\uparrow}\ket{\uparrow}   +
c_\downarrow e^{i\pi \frac{A}{2\omega_e}}  \ket{\uparrow}\ket{\downarrow} +\frac{A}{\omega_e} c_\downarrow e^{i\pi \frac{A}{2\omega_e}}  \ket{\downarrow} \ket{\uparrow}.\label{upupevolpi}
\end{eqnarray}
Comparing Eqs.~(\ref{upupevol2pi}), (\ref{upupevolpi}) we see that
the first is a separable state of the electron and  the nuclear
spin, while the second contains entanglement. This showcases the
importance of entanglement in DNP, as discussed in Section
\ref{sec:dnpQDs}, and how it manifests itself in the actual
calculated nuclear relaxation rate. Therefore, we can directly link
the sine factor in Eq.~(\ref{lambda2wn0}) to electron-nuclear
entanglement and its crucial role in the nuclear spin dynamics.

\subsection{Nuclear polarization distribution} \label{sec:nucpoldist_gen}

Once we have found the nuclear spin relaxation rate and steady
state, the nuclear spin flip rates are given by (see
Appendix~\ref{app:fliprates} for derivation and assumptions)
\begin{eqnarray}
\mathrm{w}_\pm=\gamma_n(1\pm S_{n,z}^{(\infty)})/2,\label{singlenucleusw}
\end{eqnarray}
where $\mathrm{w}_+$ ($\mathrm{w}_-$) is the rate to flip from down
(up) to up (down). More precisely, for a single nucleus we may write
\beq {dP_\uparrow\over
dt}=-\mathrm{w}_-P_\uparrow+\mathrm{w}_+P_\downarrow,
\eeq
where $P_\uparrow$ is the probability that the nucleus is aligned with the
magnetic field and $P_\downarrow=1-P_\uparrow$ is the probability
that it lies antiparallel to the magnetic field. The flip rates will generally be
different, and they will be functions of the various
parameters in the problem, including the electron Zeeman frequency $\omega_e$.

Defining the difference in the number of spins pointing up and down
as $m$, the kinetic equation for the distribution of  the net
multinuclear polarization $m/2$ is
\begin{eqnarray}
\frac{dP(m)}{dt}&=&-\sum_{\pm}\left[w_\pm(m){N\mp
m\over2}\right]P(m) \label{nucbathRE}
\\
&+&\sum_{\pm}P(m\pm2)w_\mp(m\pm2)\left[{N\pm
m\over2}+1\right],\nonumber
\end{eqnarray}
where $w_\pm(m)$ are the rates in the presence of nuclear
polarization $m/2$. These are found by implementing the Overhauser
shift, i.e., taking
\begin{eqnarray}
w_\pm(m)=\mathrm{w}_\pm(\omega_e\rightarrow\omega_e+ m A/2).
\label{fliprates}
\end{eqnarray}
In Eq.~(\ref{fliprates}) we have made the so-called `box model'
approximation, which amounts to taking all the hyperfine couplings
to be equal. This approximation is valid when the electron spin dynamics
are rapid relative to the hyperfine scale $N/{\cal A}$.\cite{Barnes_PRB11,Barnes_PRL11} This
condition is automatically satisfied whenever the driving is fast compared to the electron spin decoherence, which is the experimental regime we are considering. Note that although we are setting all the hyperfine couplings equal, we are not imposing the angular momentum symmetries associated with the box model, which would restrict the nuclear state to its original total angular momentum subspace, limiting the amount of nuclear polarization that can be generated. Here, no such limitation is present, and thus larger nuclear polarization can be generated.

The steady state corresponds to having $dP(m)/dt = 0$, which gives an
equation connecting $P(m)$ with $P(m\pm 2)$. Rearranging  terms we
can see that the following relation holds:
\begin{eqnarray}
&&\frac{N{-}
m}{2} w_+(m) P(m) {-} \left[\frac{N{+}m}{2}{+}1\right] w_-(m{+}2) P(m{+}2)
\nn\\&&=
\left[\frac{N{-}m}{2}{+}1\right] w_+(m{-}2) P(m{-}2) {-} \frac{N{+}m}{2} w_-(m) P(m),
\nonumber
\end{eqnarray}
which implies that the expression
\begin{eqnarray}
\frac{N-
m}{2} w_+(m) P(m) - \frac{N+m+2}{2} w_-(m+2) P(m+2)
\nonumber
\end{eqnarray}
is a constant. Since the full relation above is invariant with respect to rescalings of $P(m)$, we conclude that this constant is zero. We then have
\begin{eqnarray}
P(m)= \frac{N-m+2}{N+m}~\frac{w_+(m-2)}{w_-(m)} P(m-2),\label{twotermrecurrence}
\end{eqnarray}
which can be solved iteratively by setting the first nonzero entry,
$P(-N)$ to some arbitrary value and normalizing the final result: $\sum_mP(m)=1$.

\subsection{Feedback on electron spin}\label{feedbackespingeneral}

Having obtained the nuclear polarization distribution for the net
multinuclear spin system, we can calculate the updated  electron
spin vector by performing the Overhauser shift, i.e., shifting the
Zeeman frequency by $mA/2$ and averaging over $m$:
\begin{eqnarray}
\overline{S}_{e,i} = \sum_{m=-N} ^N dm P(m) S_{e,i}(\omega_e+mA/2).
\label{svavg}
\end{eqnarray}
In Eq. (\ref{svavg}) it is understood that $\overline{S}_{e,i} \equiv
\overline{S^{(\infty)}_{e,i}} $. The quantity that is usually measured experimentally in self-assembled quantum dots is
the component of $\overline{S}_e$ that is parallel (or antiparallel)
to the pulse propagation direction. Below we will derive an explicit
form for this quantity for the pulsed mode locking experiment.

\section{Application to mode locking experiment}\label{ML}

\begin{figure}[htp]
  \includegraphics[width=.9\columnwidth]{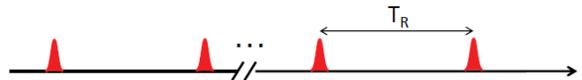}
    \caption{Pulse sequence for the mode locking experiment.}\label{fig:pulsetrain}
\end{figure}

In this section we will apply the formalism developed above to the
case where there is one fast circularly polarized pulse per period, as
depicted in Fig. \ref{fig:pulsetrain}.
There is a magnetic field pointing in the plane of the quantum dot,
perpendicular to the pulse propagation direction (the so-called
Voigt geometry). This is the experiment by Greilich et
al.,\cite{Greilich_Science07} where the periodic pumping of an ensemble
of singly charged electron dots was shown to modify the nuclear spin
environment in these dots, an effect manifested in the measurement
of the electron spin. In particular, there was a nuclear-induced
`push' of the electron Zeeman frequency towards those frequencies
that were commensurate with the pulse train period. As a result, the
ensemble obtained higher electron spin polarization than what one
would expect in the absence of the nuclear feedback. In this
experiment, resonant pulses of approximately $\pi$ area were used,
meaning that the population transfer from the electron spin state to
the optically excited trion state by the pulse was maximal. It was
later shown by Carter and collaborators\cite{Carter_PRL09} that when
detuned pulses are used instead, richer physics emerges as a result
of the interplay between coherent and incoherent pulse-induced
dynamics. In particular the detuning causes a nonzero steady state
electron spin component along the magnetic field axis, which in turn
renders the nuclear flip rates \emph{directional}, an effect absent
in the resonant case of Ref. \onlinecite{Greilich_Science07}. Here we
will treat the general case, where there is a nonzero detuning. We
will follow the steps analyzed in the subsections above for this
particular example.

\subsection{Electron spin Kraus operators}

The first step is to find  the Kraus operators
describing the electron spin evolution due to a single pulse and the
subsequent spontaneous emission. The Hamiltonian for an electron in a magnetic field along the $z$ axis,
in the absence of nuclear spin interactions and in the  presence of
a train of left-circularly polarized pulses is
\begin{eqnarray}
H_{e} &=& \omega_e \hat{S}_z + \epsilon_{\bar{T}}
|\bar{T}\rangle\langle \bar{T}|
+ \sum_k\Omega(t-k T_R) |\bar{x}\rangle\langle \bar{T}| + h.c..\nonumber\\
\end{eqnarray}
Since the g-factor of the hole along the $z$ axis is negligible,
the trion state with the opposite spin is ignored.
Note that as a result of polarization selection rules, the pulse
\emph{only} couples state $|\bar{x}\rangle$, the state with  the
electron spin pointing antiparallel to the pulse propagation
direction, to the trion state $|\bar{T}\rangle$ with angular
momentum projection $-3/2$ along the $x$ axis. In the rotating wave
approximation, the coupling to the pulse is  $\Omega(t-t_o) =
\Omega_o f(t-t_o) e^{i \omega (t-t_o)}$. The radiation field will be
included in the form of Lindblad operators. We take the pulses to be
the fastest timescale in the system, i.e., much faster than the
Zeeman precession period and the spontaneous emission timescale.
This allows us to treat the pulse as acting instantaneously on the
two-level system (comprised of $|\bar{x}\rangle$ and $|\bar{T}\rangle$) only. This is a good approximation for these types
of ultrafast experiments, where picosecond, or even subpicosecond,
pulses are used. Then we find the Kraus operators in two steps:
first we consider the coherent effects, i.e., the excitation (and
possibly stimulated emission) by the pulse, and treat the resulting
state in the three-level Hilbert space as the input to the remaining
terms describing spontaneous emission in the presence of the external magnetic field. Defining the evolution operator due to the pulse in the $\ket{x}$, $\ket{\bar x}$, $\ket{\bar T}$ basis as
\beq
U_p=\left[
\begin{array}{ccc}
1 & 0  &  0  \\
0  & u_{\bar{x}\bar{x}} & -u_{\bar{T}\bar{x}}^* \\
0 & u_{\bar{T}\bar{x}} & u_{\bar{x}\bar{x}}^*
\end{array}
\right],
\eeq
the density matrix  of the three-level system right after the pulse
is
\begin{eqnarray}
R &=& U_p R_0 U_p^\dag = \left[
\begin{array}{ccc}
\rho_{xx} & \rho_{x\bar{x}} u_{\bar{x}\bar{x}}^*  &  \rho_{x\bar{x}} u_{\bar{T}\bar{x}}^*  \\
\rho_{\bar{x}x} u_{\bar{x}\bar{x}}  & \rho_{\bar{x}\bar{x}} |u_{\bar{x}\bar{x}}|^2 & \rho_{\bar{x}\bar{x}} u_{\bar{T}\bar{x}}^*   u_{\bar{x}\bar{x}} \\
\rho_{\bar{x}x} u_{\bar{T}\bar{x}} & \rho_{\bar{x}\bar{x}}
u_{\bar{x}\bar{x}}^* u_{\bar{T}\bar{x}} & \rho_{\bar{x}\bar{x}}
|u_{\bar{T}\bar{x}}|^2
\end{array}
\right] \nonumber
\\
&\equiv &  \left[
\begin{array}{ccc}
R_{xx}^\prime & R_{x\bar{x}}^\prime & R_{x\bar{T}}^\prime \\
R_{\bar{x}x}^\prime & R_{\bar{x}\bar{x}}^\prime & R_{\bar{x}\bar{T}}^\prime  \\
R_{\bar{T}x}^\prime & R_{\bar{T}\bar{x}}^\prime &
R_{\bar{T}\bar{T}}^\prime
\end{array}
\right]. \label{rhoxbasisafterpulse}
\end{eqnarray}
Note that the expressions in Eq. (\ref{rhoxbasisafterpulse}) are in
the pulse propagation direction basis, $x$, and \emph{not} in the
energy eigenbasis. We make  this choice due to the simplicity of the
expressions coming from the optical selection rules. Subsequently
$R$ evolves under the magnetic field and the vacuum radiation field
as
\begin{eqnarray}
\dot R = i [R, \omega_e S_z] + \mathcal{L}(R).
\end{eqnarray}
Switching to the interaction picture with respect to the Zeeman
Hamiltonian, the following equations describe the evolution of the
relevant matrix elements for the  2$\times$2 spin
subspace\cite{Economou_PRB05}
\begin{eqnarray}
\dot {\widetilde{R}}_{xx} &=&  \gamma R_{\bar{T}\bar{T}} \left( 1- \cos{\omega_e t} \right), \nn\\
\dot {\widetilde{R}}_{\bar{x}\bar{x}} &=& \gamma R_{\bar{T}\bar{T}} \left( 1+ \cos{\omega_e t} \right), \nn\\
\dot {\widetilde{R}}_{x\bar{x}} &=& i \gamma R_{\bar{T}\bar{T}} \sin{\omega_e t}, \nn\\
\dot {R}_{\bar{T}\bar{T}} &=& -2 \gamma R_{\bar{T}\bar{T}} ,
\label{decaywsgc}
\end{eqnarray}
where $\widetilde{R}$ is the density matrix in the interaction
picture. Eqs. (\ref{decaywsgc}) include the so-called spontaneously
generated coherence effect, \cite{Javanainen_EL92,Economou_PRB05,Dutt_PRL05}
which results from the fact that due to  polarization selection
rules, spontaneous emission couples state $|\bar{T}\rangle$ to
$|\bar{x}\rangle$ only [although spontaneous emission \emph{together} with precession
leads to some population decaying to $|x\rangle$ as well, as seen in the topmost equation of Eqs. (\ref{decaywsgc})].
 This effect is significant when the Zeeman frequency, $\omega_e$, is
smaller or comparable to the relaxation rate, $\gamma$. From the last equation we readily obtain
\begin{eqnarray}
R_{\bar{T}\bar{T}}=R_{\bar{T}\bar{T}}^\prime e^{-2\gamma t},
\end{eqnarray}
which then allows us to find the matrix elements in the spin
subspace by a simple integration. Doing that and taking the limit
$t\gg \gamma ^{-1}$ we find
\begin{eqnarray}
\widetilde{R}_{xx}&=& R_{xx}^\prime +\frac{\omega_e^2}{2 \left(4 \gamma^2+\omega_e^2\right)} R_{\bar{T}\bar{T}}^\prime, \label{RxxafterSE}
\\
\widetilde{R}_{\bar{x}\bar{x}}&=&
R_{\bar{x}\bar{x}}^\prime+\left[\frac{2 \gamma^2} {4
\gamma^2+\omega_e ^2}+\frac{1}{2} \right] R_{\bar{T}\bar{T}}^\prime,
\label{RxbarxbarafterSE}
\\
\widetilde{R}_{x\bar{x}}&=& R_{x\bar{x}}^\prime +i\frac{\gamma
\omega_e} {4 \gamma^2+\omega_e^2} R_{\bar{T}\bar{T}}^\prime .
\label{RxxbarafterSE}
\end{eqnarray}
Notice that the degree of spin polarization depends on the ratio
$\omega_e/\gamma$, as was discussed in detail in Ref. \onlinecite{Economou_PRB05}.
In our previous work,\cite{Barnes_PRL11} we considered the high
magnetic field limit, i.e., we assumed $\omega_e/\gamma \gg 1$. In
that limit Eqs.~(\ref{RxxafterSE})-(\ref{RxxbarafterSE}) above
simplify and the  coefficients of $R_{\bar{T}\bar{T}}^\prime$ are
1/2 for (\ref{RxxafterSE}), (\ref{RxbarxbarafterSE}) and zero for
(\ref{RxxbarafterSE}). Here we relax that assumption to account for
low B-fields.

Combining Eqs.~(\ref{rhoxbasisafterpulse}) and (\ref{RxxafterSE})-(\ref{RxxbarafterSE}) we
obtain for the spin density matrix  in the lab frame after the pulse and spontaneous emission
\begin{eqnarray}
\rho_{xx}^\prime &=& \rho_{xx} + \frac{\omega_e^2}{2 \left(4
\gamma^2+\omega_e^2\right)} |u_{\bar{T}\bar{x}}|^2
\rho_{\bar{x}\bar{x}}, \label{rprimexx}
\\
\rho_{\bar{x}\bar{x}}^\prime &=& \rho_{\bar{x}\bar{x}}
|u_{\bar{x}\bar{x}}|^2 +\left[\frac{2 \gamma^2}{4 \gamma^2+\omega_e
^2}+\frac{1}{2} \right]  |u_{\bar{T}\bar{x}}|^2
\rho_{\bar{x}\bar{x}}, \label{rprimebxbx}
\\
\rho_{x\bar{x}}^\prime &=& \rho_{x\bar{x}}
u_{\bar{x}\bar{x}}^* + i \frac{\gamma\omega_e }{ \left(4 \gamma^2+\omega_e
^2\right)}  |u_{\bar{T}\bar{x}}|^2 \rho_{\bar{x}\bar{x}},\label{rprimexbx}
\end{eqnarray}
where we have used that $\tilde\rho_{ij}^\prime=\rho_{ij}^\prime$.
Using the unitarity of $U_p$, i.e., setting
$|u_{\bar{T}\bar{x}}|^2=1-|u_{\bar{x}\bar{x}}|^2$ and  by inspection
of Eqs. (\ref{rprimexx})-(\ref{rprimexbx}) we obtain the Kraus
operators in the lab frame ($x$ basis):
\begin{eqnarray}
E_1 &=&
\left[
\begin{array}{cc}
1 & 0  \\
0 & q
\end{array}
\right],\nonumber
\\
E_2&=&
\left[
\begin{array}{cc}
0 & \;\;\;a_1  \\
0 & -a_2
\end{array}
\right],\nonumber
\\
E_3 &=&
\left[
\begin{array}{cc}
0 & 0  \\
0 & \kappa
\end{array}
\right],\label{krausops}
\end{eqnarray}
where
\begin{eqnarray}
q &=& u_{\bar{x}\bar{x}} \equiv q_o e^{i\phi},
\\
a_1&=&\omega_e\sqrt{\frac{(1-q_o^2)}{2(4\gamma^2+\omega_e^2)}},
\\
a_2 &=& i \gamma\sqrt{2} \sqrt{\frac{1-q_o^2}{4\gamma^2+\omega_e^2}},
\\
\kappa &=& \sqrt{1-q_o^2-a_1^2-|a_2|^2}.
\end{eqnarray}
In the limit $\omega_e \gg \gamma$, where the spontaneously generated coherence effect is negligible, we have $a_2 \rightarrow 0  $ and
$a_1,\kappa \rightarrow \sqrt{(1-q_o^2)/2}$, so that we recover the
Kraus  operators from Ref. \onlinecite{Barnes_PRL11}.

Here, it is important to clarify a point of potential confusion: In deriving the above Kraus operators, we
assumed $\gamma T_R\gg1$, so that the trion decays completely back to the electron
spin subspace within a single period. Therefore, setting $\gamma=0$ in the Kraus operators
\emph{does not} correspond to the absence of spontaneous emission, but instead it corresponds
to neglecting spontaneously generated coherence, i.e., to the case of equal decay to both electron spin
states. In what follows, we will see that the ratio $\gamma/\omega_e$
plays an important role in the generation of DNP.

The matrix elements of $U_p$ are functions of the pulse-system
parameters, namely the Rabi frequency, the detuning, the bandwidth
and the pulse shape. For example, in the case where the pulse has the shape of
the hyperbolic secant, which is analytically solvable, the matrix element
$u_{\bar{x}\bar{x}}$ has the explicit form\cite{Economou_PRB06}
\begin{eqnarray}
u_{\bar{x}\bar{x}} = F(a,-a,c^\ast,1) =\frac{\Gamma(c)^2}{\Gamma(c-a)\Gamma(c+a)},
\end{eqnarray}
where $F$ is Gauss's hypergeometric function, $\Gamma$ is the gamma function, and $a=\Omega_o/\sigma$ and $c=1/2(1+i\Delta/\sigma)$, with
$\Omega_o$, $\Delta$, and $\sigma$ denoting the Rabi frequency, pulse detuning, and bandwidth, respectively.
 In this case, the Kraus parameter $q_o$ can be expressed as
\beq
q_o=|u_{\bar{x}\bar{x}}|=\sqrt{1-\sin^2(\pi\Omega_o/\sigma)\hbox{sech}^2(\pi\Delta/2\sigma)}.
\eeq
In principle, $q_o$ and $\phi$ can be computed for any pulse shape; thus we will continue to
express results in terms of these parameters for the sake of generality.

The quantities $q_o$ and $\phi$ are two of the key parameters of the theory. Physically, $1-q_o^2$ is the fraction of population
that moves from the electron spin state $\ket{\bar x}$ to the trion state $\ket{\bar T}$, while
$\phi$ is the angle about the $x$ axis by which the pulse rotates the electron spin, with its sign
coinciding with that of the detuning, $\Delta$.

\subsection{Electron and nuclear spin steady states and nuclear relaxation rate}
Transforming the Kraus operators of Eq.~(\ref{krausops}) to the $z$ basis and using Eqs.~(\ref{K3})-(\ref{Y3}), we find
\bea
K_e&=&\left[\begin{matrix} a_1^2 & -ia_1a_2 & 0 \end{matrix}\right],\nn\\
Y_e&=&\left[\begin{matrix} 1-a_1^2 & 0 & 0\cr ia_1a_2 & q_o\cos\phi & -q_o\sin\phi \cr 0 & q_o\sin\phi & q_o\cos\phi \end{matrix}\right].
\eea
Combining these results with the evolution operator describing precession between pulses,
\beq
Y_{pr}=\left[\begin{matrix} \cos(\omega_eT_R) & -\sin(\omega_eT_R) & 0 \cr \sin(\omega_eT_R) & \cos(\omega_eT_R) & 0 \cr 0 & 0 & 1 \end{matrix}\right],\label{Ypr}
\eeq
we obtain the electron steady state right after each pulse from the formula, $S_e^{(\infty)}=(1-Y_eY_{pr})^{-1}K_e$, with
\begin{widetext}
\bea
S_{e,x}^{(\infty)}\!\!\!&{=}&\!\!\!\frac{a_1 \left(a_1 q_o \left(q_o-\cos\phi\right) \cos \left(\omega _e T_R\right)-i a_2 \left(q_o \cos\phi-1\right) \sin \left(\omega _e
   T_R\right)-a_1 q_o\cos\phi+a_1\right)}{\left(a_1^2+q_o^2-1\right) \cos \left(\omega _e T_R\right)-a_1 q_o \cos \phi  \left[i a_2
   \sin \left(\omega _e T_R\right)+a_1 \cos \left(\omega _e T_R\right)+a_1\right]+i a_1 a_2 \sin \left(\omega _e T_R\right)+\left(a_1^2-1\right)
   q_o^2+1},\nn\\
S_{e,y}^{(\infty)}\!\!\!&{=}&\!\!\!\frac{a_1 \left(a_1 q_o \left(\cos\phi-q_o\right) \sin \left(\omega _e T_R\right)-i a_2 \left(q_o \cos\phi-1\right) \left(\cos
   \left(\omega _e T_R\right)-1\right)\right)}{\left(a_1^2+q_o^2-1\right) \cos \left(\omega _e T_R\right)-a_1 q_o \cos\phi \left[i a_2 \sin
   \left(\omega _e T_R\right)+a_1 \cos \left(\omega _e T_R\right)+a_1\right]+i a_1 a_2 \sin \left(\omega _e T_R\right)+\left(a_1^2-1\right)
   q_o^2+1},\nn\\
S_{e,z}^{(\infty)}\!\!\!&{=}&\!\!\!\frac{a_1 q_o \sin\phi \left(a_1 \sin \left(\omega _e T_R\right)-i a_2 \left(\cos \left(\omega _e
   T_R\right)-1\right)\right)}{\left(a_1^2+q_o^2-1\right) \cos \left(\omega _e T_R\right)-a_1 q_o \cos\phi \left[i a_2 \sin \left(\omega _e
   T_R\right)+a_1 \cos \left(\omega _e T_R\right)+a_1\right]+i a_1 a_2 \sin \left(\omega _e T_R\right)+\left(a_1^2-1\right) q_o^2+1}.\nn\\&&\label{espinss}
\eea
\end{widetext}
Note that the steady state undergoes Larmor precession during each period and that the state at any point during the period can be obtained by evolving the above expressions using Eq.~(\ref{Ypr}).

\begin{figure}[htp]
  \includegraphics[width=.85\columnwidth]{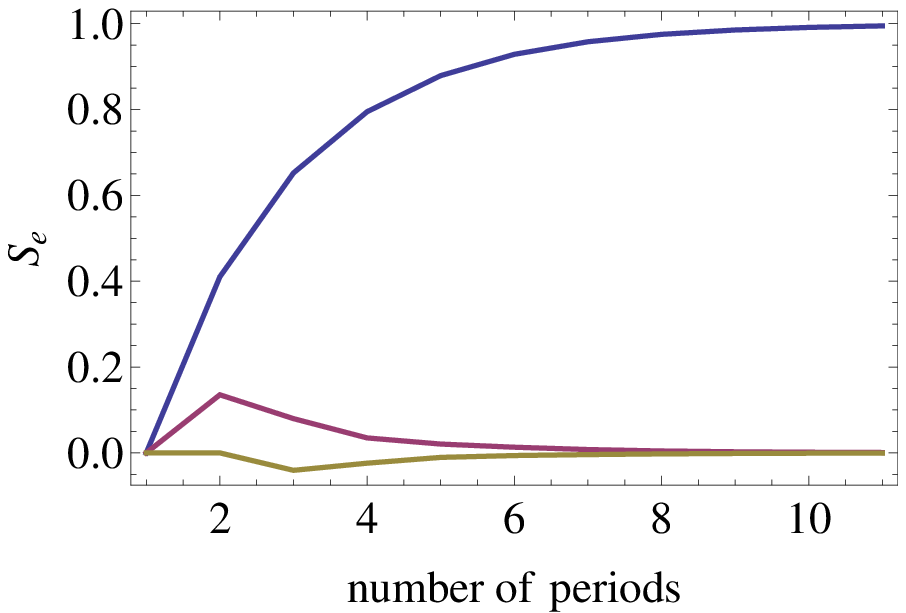}
    \includegraphics[width=.85\columnwidth]{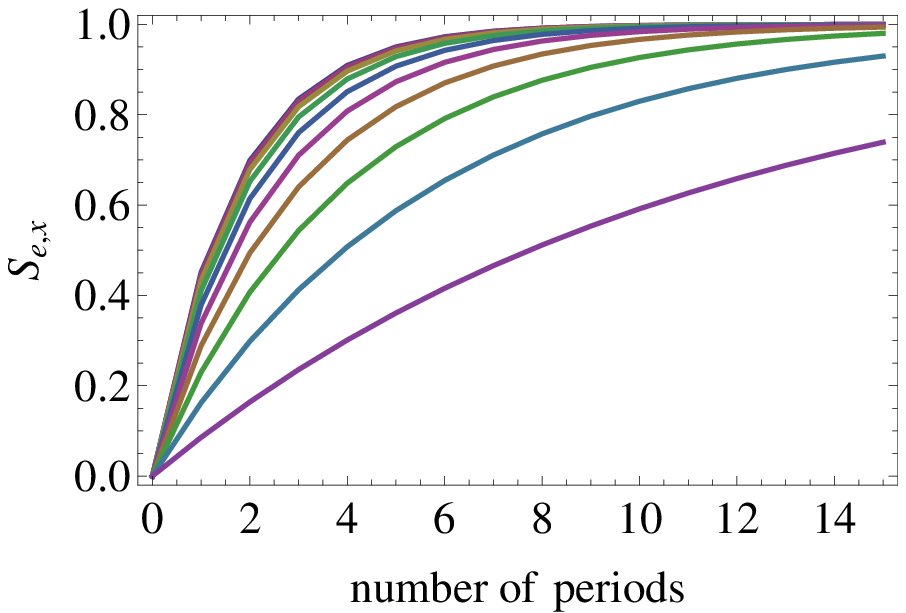}
\caption{Coarse-grained electron spin evolution for $T_R=13.2$ns,
$\gamma=0.5$GHz,  $\phi=-\pi/2$. Top: all three components  (top to
bottom: $S_{e,x},S_{e,y},S_{e,z}$ ) for  $q_o=0.3$ and bottom:
$S_{e,x}$  for $q_o=0,0.1,...,0.9$ (top to bottom). }\label{fig:coarsegraines}
\end{figure}

At this point it is useful to verify the separation of timescales
necessary for the validity of the Markovian approximation discussed
in Section \ref{sec:includesinglesnspin_gen}. In particular, we want
to show that the time it takes for the electron spin to reach this
steady state, which we define as $\tau_e$, is small compared to the
typical decoherence time. As in the case of the nuclear spin
(discussed in Section \ref{sec:includesinglesnspin_gen}), we can
obtain $\tau_e$ from the eigenvalues of $\mathbbm{1}-\mathcal{Y}_e$.
In the special case of resonant $\pi$ pulses, $q_o=0$, we obtain the
analytical expression:
\begin{eqnarray}
\tau_e=\frac{2 T_R \left(\omega _e^2+4 \gamma^2\right)}{2 \left(\omega _e^2{+}4
   \gamma^2\right){-}\left(\omega _e^2{+}8
   \gamma^2\right) \cos \left(\omega _e T_R\right){+}2 \gamma \omega _e \sin \left(\omega _e T_R\right)}.\nonumber
\end{eqnarray}
From this expression, it is clear that the slowest relaxation times
occur when the frequency is commensurate with the pulse train
period. In that case we  obtain $\tau_e \rightarrow (2+8 \gamma^2/\omega
_e^2)T_R$. We have also checked that a similar timescale holds for
other values of $q_o\lesssim 0.5$. In Fig. \ref{fig:coarsegraines}
we plot the spin vector components as functions of the coarse-grained
time for $q_0=0.3$, and also show the trend for the coarse-grained
evolution of the electron spin $x$ component as a function
of $q_o$. From these figures, it is evident that the electron spin
reaches its steady state after only a few periods.
For typical pulse periods $T_R\sim 10$ns, we have $\tau_e
\lesssim 100$ns, which is well below typical decoherence times of
several microseconds, justifying  the use of a Markovian approach. For values
of $q_o\gtrsim0.5$, it is apparent from the lower panel of Fig. \ref{fig:coarsegraines}
that the electron spin reaches its steady state sufficiently slowly that
the validity of the Markovian approach is questionable. This highlights
the intrinsic connection between the smallness of $q_o$ and Markovianity, which
was pointed out in Ref. \onlinecite{Barnes_PRL11}. For the numerical results
we present below, we use $q_o=0.3$, a value which is both well within the Markovian
regime and also large enough that coherent effects due to the pulses are significant.

\subsection{Nuclear steady state and relaxation rate} \label{sec:nssrelaxn_ML}

\begin{figure}[htp]
  \;\includegraphics[width=.45\columnwidth]{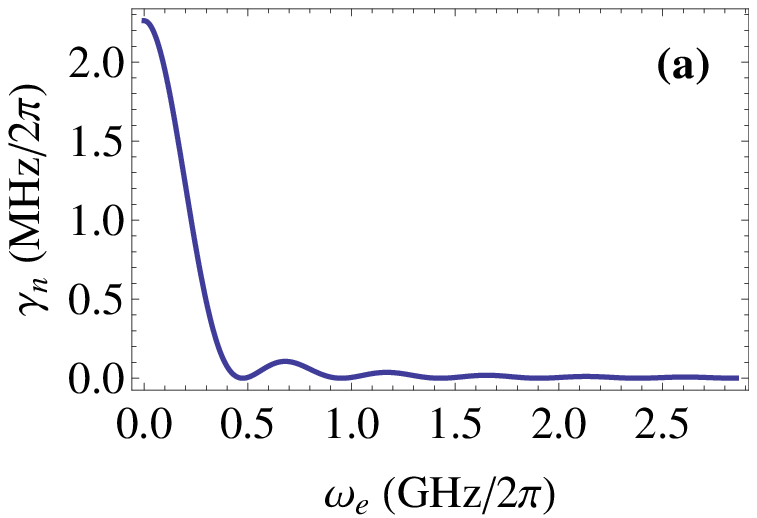}\;\;
  \includegraphics[width=.46\columnwidth]{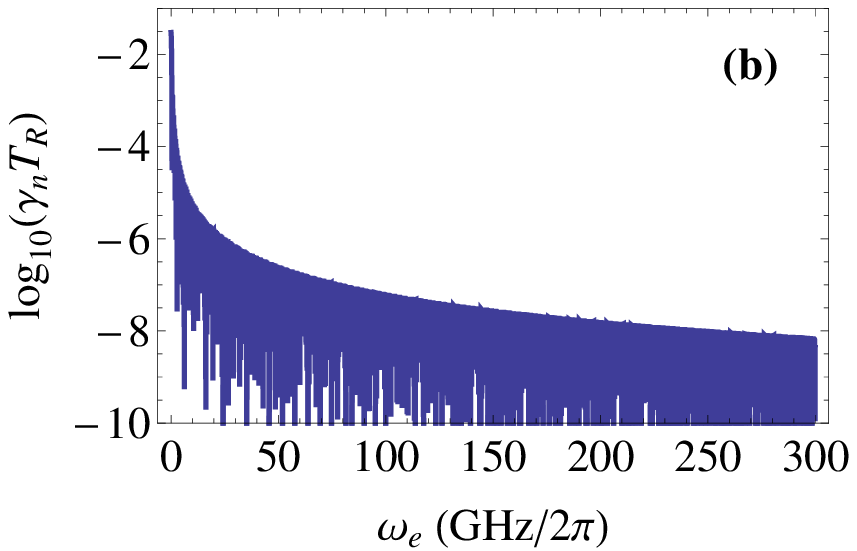}
  \includegraphics[width=.48\columnwidth]{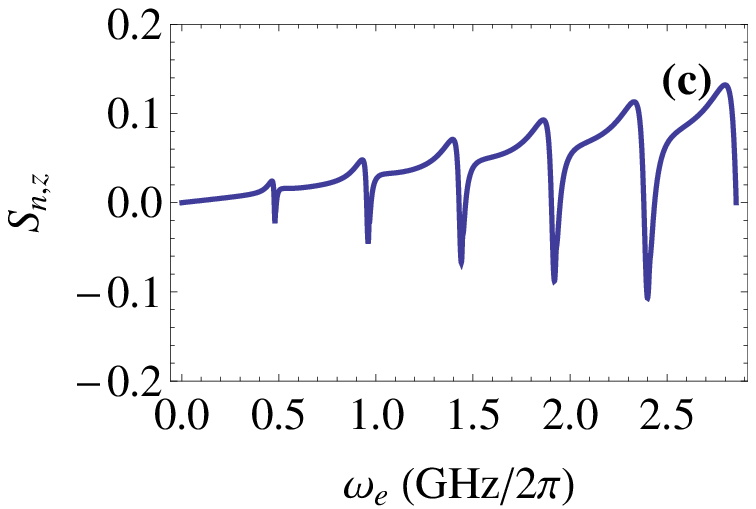}
  \includegraphics[width=.49\columnwidth]{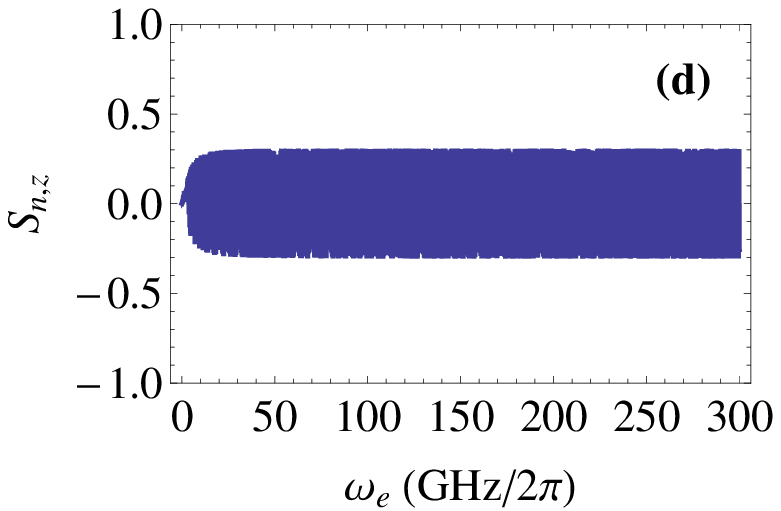}\;\;
  \;\;\;\;\;\includegraphics[width=.43\columnwidth]{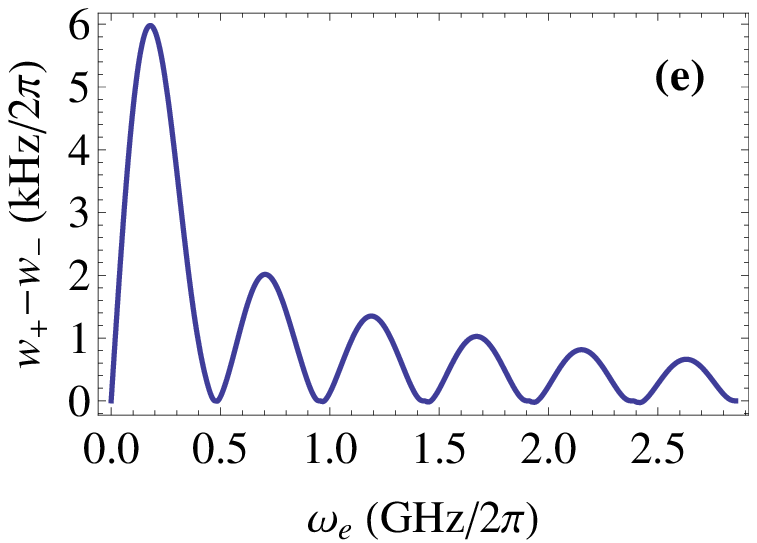}
  \;\includegraphics[width=.48\columnwidth]{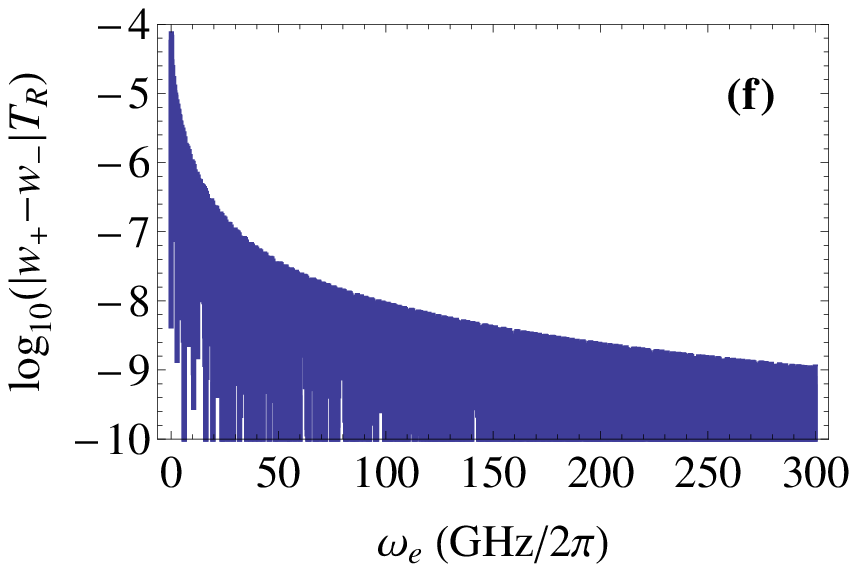}
  \caption{(a,b) Nuclear relaxation rate, (c,d) steady state, and (e,f) difference
of single-nucleus flip rates versus electron Zeeman energy for
$T_R=13.2$ns, $NA=12.5$GHz, $N=3000$, $\gamma=0.5$GHz, $q_0=0.3$,
$\phi=-\pi/2$, $\omega_n=0$. The expression (GHz/$2\pi$) in the units of $\omega_e$ is equivalent to radians/nanosecond.}\label{fig:relaxfliprates}
\end{figure}

Given  expressions (\ref{espinss}) for the electron steady state, we
can compute the nuclear spin steady state and relaxation
rate from Eqs.~(\ref{snzwn0})-(\ref{lambda2wn0}). The nuclear relaxation
rate $\gamma_n=\lambda_2^*/T_R$ is shown in Figs.~\ref{fig:relaxfliprates}(a,b) as a
function of the electron Zeeman energy. It is apparent from the
figures that this rate becomes larger as the Zeeman energy
decreases. This trend is due to the fact that the electron spin
flips more easily with nuclear spins when its Zeeman energy is
smaller, leading to faster relaxation.
Figs.~\ref{fig:relaxfliprates}(c-f) reveal that this feature of
$\gamma_n$ carries over to the difference of the single-nucleus spin
flip rates, $\mathrm{w}_\pm$, even though the nuclear steady state
is larger at higher magnetic fields.
As we will see in the next subsection, this difference in flip rates
leads to  a nonzero average nuclear spin polarization at lower
magnetic fields. Note that in Fig.~\ref{fig:relaxfliprates}, we have
taken the rotation angle to be $\phi=-\pi/2$. Reversing the sign of
$\phi$ changes the sign of $S_{e,z}^{(\infty)}$ and hence the sign
of $w_+-w_-$. Thus the sign of the nuclear spin polarization depends
directly on the sign of $\phi$.

It is also apparent from Fig.~\ref{fig:relaxfliprates}
that $\gamma_n$ and hence the flip rate difference periodically go to zero. This
behavior stems directly from the sine factor in Eq. (\ref{lambda2wn0}), which vanishes
when the Zeeman precession period is commensurate with the pulse period. When this condition is satisfied,
the electron spin is polarized along the $x$ direction when the pulse arrives and is thus unaffected
by the pulses, removing the mechanism through which the nuclear spin attains its steady state
and leading to $\gamma_n=0$. This feature of $\gamma_n$ plays a central role in the frequency focusing effect,
as is explained in the next subsection.

It should be mentioned that since the results shown in Fig.~\ref{fig:relaxfliprates} were obtained using a
perturbation theory that assumes $A/\omega_e\gg1$, we cannot trust the results for values of $\omega_e$ very close to zero. For the
parameters used in the plots, this implies that perturbation theory is valid for $\omega_e\gg0.026$ radians/ns. This condition should
also be kept in mind below when we include the Overhauser shift to obtain the effective electron Zeeman frequency, which must
satisfy the same condition. For the external magnetic fields we consider and Overhauser shifts we calculate, this condition is satisfied
for all the results we obtain in the paper.

\subsection{Incorporating multinuclear effects: self-consistent Overhauser shift}\label{multinOVh}

\begin{figure*}[t!]
  \includegraphics[width=\textwidth]{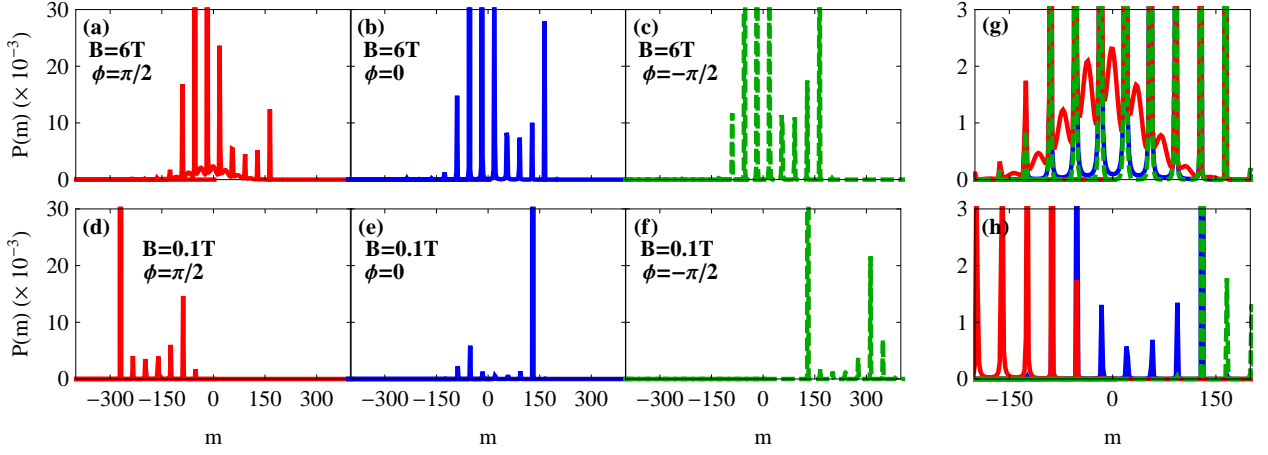}
  \caption{(a-f) Nuclear spin polarization distribution for $T_R=13.2$ns, $NA=12.5$GHz, $N=3000$, $\gamma=0.5$GHz, $q_o=0.3$, $\omega_n=0$ for various magnetic fields and rotation angles. Note that several of the peaks extend well beyond the vertical range displayed. (g) Zoom-in of (a,b,c). (h) Zoom-in of (d,e,f).}\label{fig:Pofm}
\end{figure*}
Using the flip rates for a single nuclear spin obtained in the
previous subsection, we shift the  electron Zeeman frequency by the
Overhauser field as described by Eq.~(\ref{fliprates}) to obtain
flip rates that take into account the effect of the
full nuclear
spin bath. We then feed these flip rates into the recurrence
relation, Eq.~(\ref{twotermrecurrence}), that defines the steady
state solution of the nuclear polarization rate equation
(\ref{nucbathRE}). We solve this recurrence relation numerically for
small ($B=0.1$T) and large ($B=6$T) values of the magnetic field and
for three different values of the pulse rotation angle $\phi$
($0,\pm\pi/2$). We choose the electron g-factor to be such that
$\omega_e/B=7.14$GHz/T, so that the two values of $B$ we consider
correspond to $\omega_e=4.5$ radians/ns and
$\omega_e=269.2$ radians/ns. The results are shown in
Fig.~\ref{fig:Pofm}.

Each panel of the figure clearly shows that the polarization
distribution generally exhibits a series of equally spaced peaks.
The spacing is given by $\Delta m=4/(AT_R)$ and is a direct
manifestation of the frequency focusing phenomenon described in
Ref.~\onlinecite{Greilich_Science07} in which the nuclear polarization
builds up in such a way as to shift the electron Zeeman frequency to
values commensurate with the pulse frequency. In particular, the
polarization peaks are located at values of $m$ such that
$(\omega_e+Am/2)T_R$ is an integer multiple of $2\pi$. The physics
leading to this effect is as follows: an electron spin with Zeeman
frequency $\omega_e\neq 2\pi n/T_R$ will undergo dynamics that cause
the nuclear spins to flip, see Eq.~(\ref{lambda2wn0}).
Through this process, the Overhauser shift will
alter the dynamics of the electron spin itself, until the shifted
electron Zeeman frequency satisfies the relation $\omega_e = 2\pi
n/T_R$, at which point the process stops since the nuclear spin flip
rates vanish at these values of $\omega_e$. Therefore, there is a
tendency for the system to synchronize with the pulses, leading to
the sharp, equally spaced peaks of Fig.~\ref{fig:Pofm}. This comb-like
structure can in fact be derived analytically by taking the continuum
limit of the kinetic equation, as shown in Appendix \ref{app:continuum}. While the variance of the full distribution is comparable to that of a thermal state, each peak is substantially narrower, and this should lead to longer coherence times for the electron spin.

This focusing effect arises directly from the sine factor in
Eq. (\ref{lambda2wn0}) and is independent of whether $\mathrm{w}_+$ is larger
or smaller than $\mathrm{w}_-$. However,  this effect can either be enhanced or reduced
depending on the behavior of $\mathrm{w}_+-\mathrm{w}_-$ in the
vicinity of the synchronization points: \bea
\mathrm{w}_+-\mathrm{w}_-&=&\gamma_n(2\pi n/T_R+\delta\omega_e)S_{n,z}^{(\infty)}(2\pi n/T_R+\delta\omega_e)\nn\\
&\approx
&\frac{A^2T_R^4q_o\sin\phi}{8\pi^2(1+q_o^2-2q_o\cos\phi)}\delta\omega_e^3.
\eea Consider first the case $\phi<0$. In this case, when
$\delta\omega_e>0$, we have $\mathrm{w}_+<\mathrm{w}_-$, so that
there is a tendency to generate  negative nuclear polarization. This
negative polarization will shift $\omega_e$ toward smaller values
via the Overhauser shift.
On the other hand, when $\delta\omega_e<0$, we have
$\mathrm{w}_+>\mathrm{w}_-$, and positive polarization is produced,
shifting $\omega_e$ toward larger values. Thus we see that when
$\phi<0$, the commensurate values $\omega_e=2\pi n/T_R$ are stable
fixed points, and nuclear polarization forms in such a way as to
drive the effective electron Zeeman frequency toward these values,
further enhancing the sharp, evenly spaced peaks in
Fig.~\ref{fig:Pofm}. This enhancement is particularly evident in the
feedback effect on the electron spin that will be examined in the
next subsection.

When $\phi>0$, we have the reverse situation, where now positive
deviations $\delta\omega_e>0$ lead to positive polarization and
negative  deviations to negative polarization. Therefore, in this
case, the commensurate values $\omega_e=2\pi n/T_R$ become unstable
fixed points, with a nuclear polarization-driven repulsion of the
effective $\omega_e$ away from these points. This
`anti-synchronization' effect is evident in
Figs.~\ref{fig:Pofm}(a,g), where the curve corresponding to
$\phi=\pi/2$ exhibits additional, broad peaks centered in between
the narrow peaks. This effect was first studied theoretically and
experimentally in Ref. \onlinecite{Carter_PRL09}. In this regime, one might be tempted to say that there exists a stable stationary state between two adjacent, repulsive commensurate points, however such a state would only be approximately stationary. This is because the nuclear spin flip rates are nonzero for all values of the electron Zeeman frequency between the two commensurate points, implying that the state continues to evolve. Since this evolution is constrained by the two repulsive fixed points, we envision the state as going back and forth between them, such that the system spends more time at the half-commensurate points on average, leading to an approximate stationary state there. This is to be contrasted with the truly stable fixed points that occur at the commensurate values when $\phi<0$. At these points, the nuclear flip rates are precisely zero, signifying a real stationary state. Both synchronization and
anti-synchronization peaks appear in Fig.~\ref{fig:Pofm}(a) since
the $\phi$-independent focusing effect in which nuclear spin
fluctuations randomly shift the electron Zeeman frequency to
commensurate values is still present.

\begin{figure}
  \includegraphics[width=0.8\columnwidth]{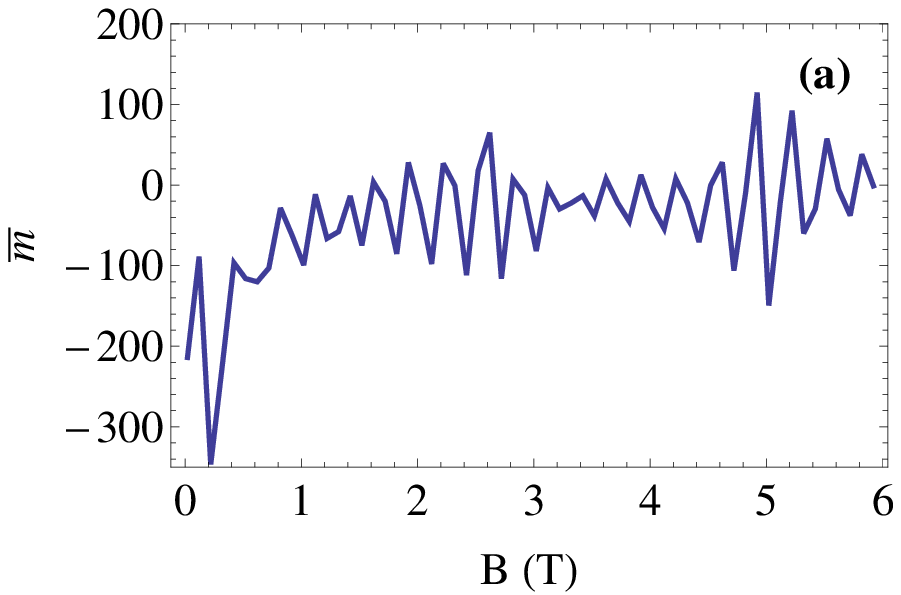}
  \includegraphics[width=0.8\columnwidth]{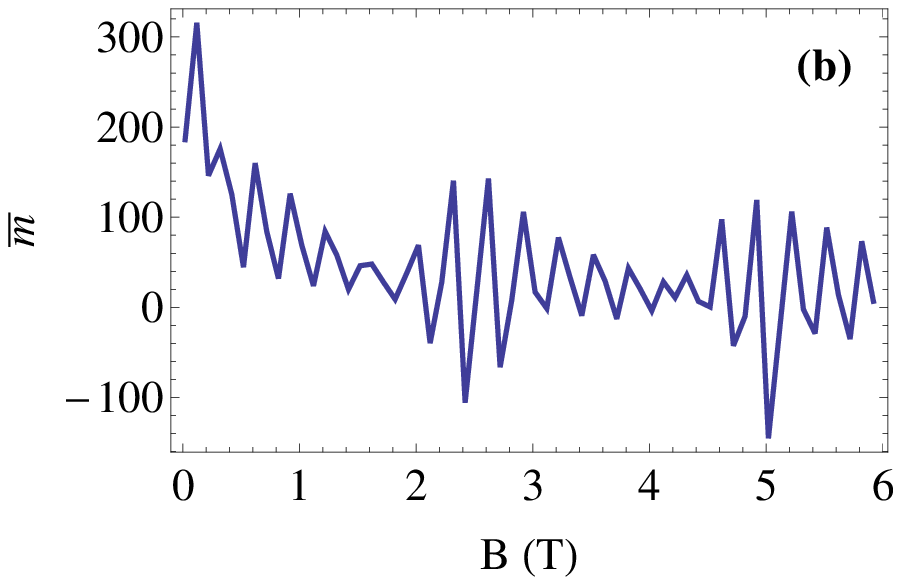}
  \caption{Average nuclear spin polarization versus magnetic field for $T_R=13.2$ns, $NA=12.5$GHz, $N=3000$, $\gamma=0.5$GHz, $q_o=0.3$, $\omega_n=0$, and (a) $\phi=\pi/2$, (b) $\phi=-\pi/2$.}\label{fig:mave}
\end{figure}

It is also apparent from Fig.~\ref{fig:Pofm} that the polarization
distribution $P(m)$ is centered around nonzero  polarizations when
$B=0.1$T and $\phi$ is nonzero. Fig.~\ref{fig:Pofm} further reveals that the sign of the net
polarization that occurs at low magnetic fields is  opposite to the
sign of $\phi$, as was anticipated in the previous subsection. The
values $\phi=\pm\pi/2$ chosen for the figure give rise to maximal
values for the polarization; the magnitude of the polarization
increases steadily up to $\phi=\pm\pi/2$, and beyond these values,
the polarization steadily decreases, returning to values close to
zero at $\phi=\pm\pi$. In Fig.~\ref{fig:Pofm}, it is also clear that
the net polarization is significantly reduced at large magnetic
fields even for $\phi=\pm\pi/2$. This trend is more explicit in
Fig.~\ref{fig:mave}, where we plot the average nuclear polarization,
$\bar m=\sum_m mP(m)$, as a function of magnetic field.
This figure shows that the DNP is largest at low magnetic fields and
is suppressed at high magnetic fields. This behavior originates from
the phenomenon  of spontaneously generated coherence, which will be
discussed in detail in Section \ref{sec:sgc}.

\subsection{DNP feedback on electron spin}\label{modelockingfeedback}

\begin{figure}
  \includegraphics[width=\columnwidth]{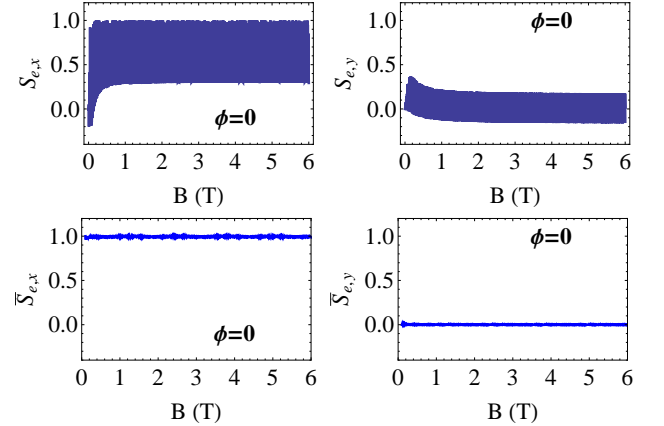}
  \caption{{\it Effect of nuclear feedback:} Electron spin steady state versus magnetic field for $T_R=13.2$ns, $NA=12.5$GHz, $N=3000$, $\gamma=0.5$GHz, $q_o=0.3$, $\omega_n=0$, $\phi=0$. Upper panels: without nuclear feedback (Eq.~(\ref{espinss})); lower panels: with nuclear feedback (Eq.~(\ref{svavg})).}\label{fig:essphi0}
\end{figure}
\begin{figure*}
  \includegraphics[width=0.8\textwidth]{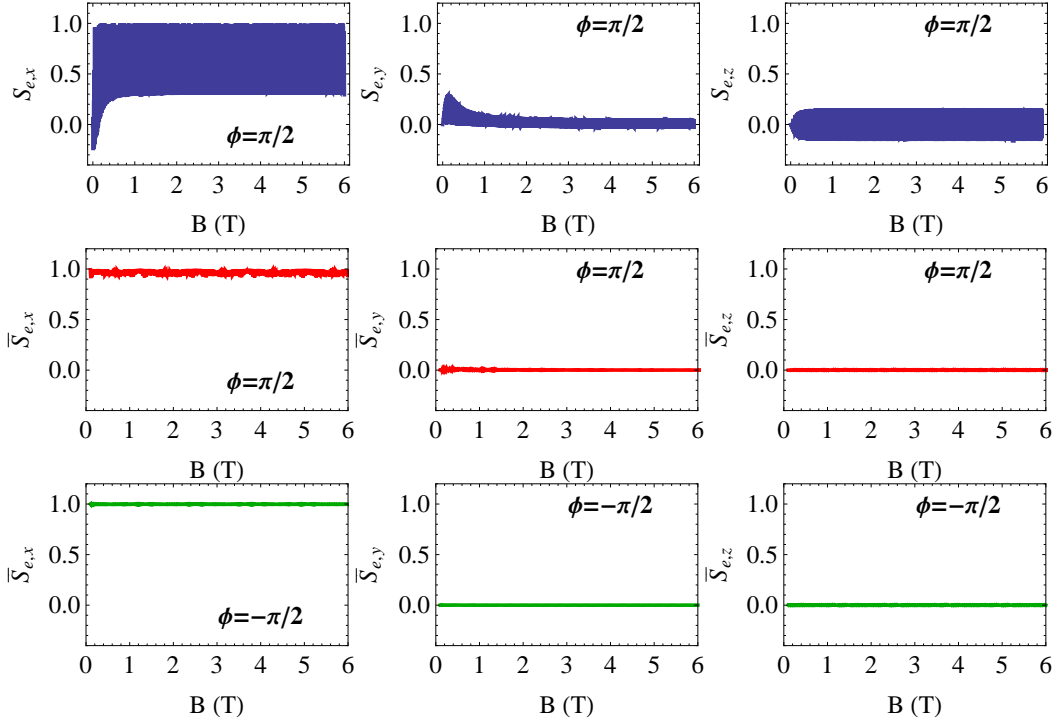}
  \caption{{\it Effect of nuclear feedback:} Electron spin steady state versus magnetic field for $T_R=13.2$ns, $NA=12.5$GHz, $N=3000$, $\gamma=0.5$GHz, $q_o=0.3$, $\omega_n=0$. Top row: without nuclear feedback (Eq.~(\ref{espinss})) for $\phi=\pi/2$ (the case with $\phi=-\pi/2$ is essentially the same); middle row: with nuclear feedback (Eq.~(\ref{svavg})) for $\phi=\pi/2$; bottom row: with nuclear feedback (Eq.~(\ref{svavg})) for $\phi=-\pi/2$.}\label{fig:essphinot0}
\end{figure*}
\begin{figure*}
  \includegraphics[width=0.8\textwidth]{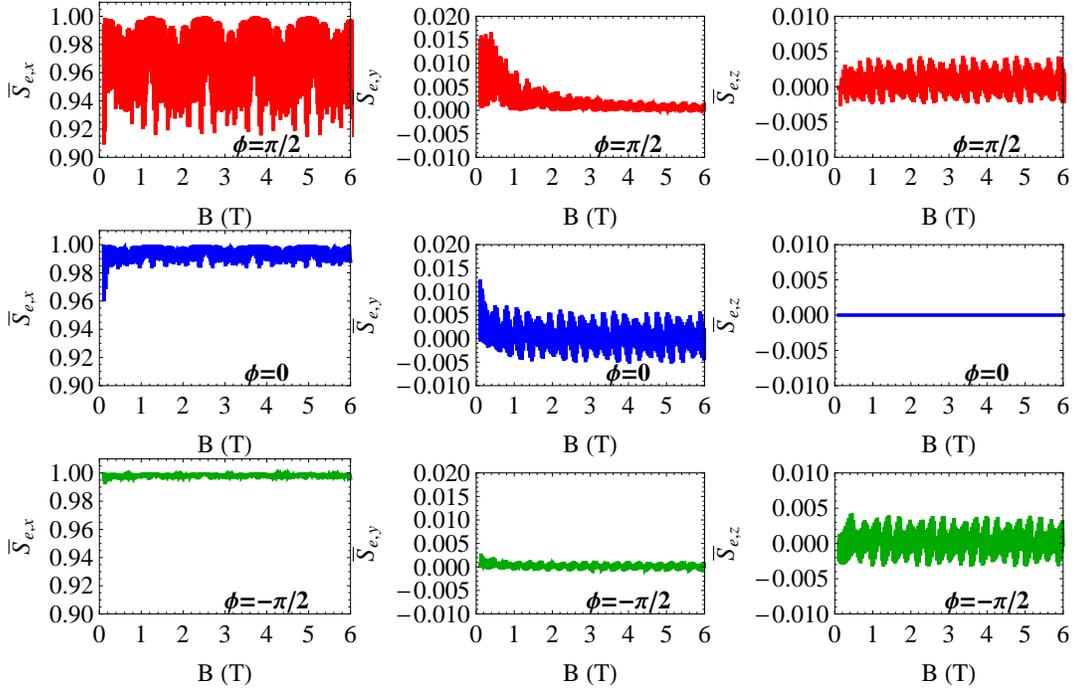}
  \caption{{\it Effect of nuclear feedback:} Zooming in on electron spin steady state versus magnetic field including nuclear feedback for $T_R=13.2$ns, $NA=12.5$GHz, $N=3000$, $\gamma=0.5$GHz, $q_o=0.3$, $\omega_n=0$ for the three cases, $\phi=0,\pm\pi/2$. In the case $\phi=\pi/2$, the anti-focusing effect is evident in the $x$-component of the electron steady state, while the case $\phi=\pi/2$ gives stronger focusing compared to the $\phi=0$ case.}\label{fig:essphizoomin}
\end{figure*}
Having calculated the nuclear spin distribution, we proceed to find
the feedback on the electron steady state. The total nuclear
polarization shifts the  Zeeman frequency of the electron spin, and
the features of the nuclear distribution described in the previous
subsection are therefore anticipated to appear in the final electron
spin steady state. Using the expressions for the electron spin
components, Eqs.~(\ref{espinss}), we average $\omega_e$ over the
distribution $P(m)$, as explained in subsection
\ref{feedbackespingeneral}.  The resulting electron spin vector is
shown without and with feedback for $\phi=0$  in
Fig.~\ref{fig:essphi0}, and for  $\phi=\pm\pi/2$ in
Fig.~\ref{fig:essphinot0}. From these figures it is evident that the SV components oscillate rapidly as functions of magnetic field, and that the amplitude of these oscillations is dramatically reduced when nuclear feedback is included. Furthermore, we see that the $x$ component of the electron spin vector tends to 1, while the $y,z$ components tend to 0. This shows that the nuclear polarization is built up in such a way that it synchronizes the electron spin with the pulse so that its steady state at the time of the pulse is now fully polarized along $x$. Note also the effect of synchronization and anti-synchronization (discussed in the previous subsection) in the
spin vector components. The nuclear feedback in both cases
focuses the electron Zeeman splitting through synchronization, but
in the case of $\phi=\pi/2$ the effect is weaker. This can be seen
more clearly in Fig.~\ref{fig:essphizoomin}, where we have zoomed in
on the spin components for all three cases, $\phi=0,\pm\pi/2$,
revealing an increasing synchronization effect as we move from
positive to negative $\phi$. In particular, we see that in the case of positive $\phi$ the amplitude of oscillation of the $S_{e,x}$ component is significantly larger as compared to zero and negative $\phi$'s.

\subsection{Effects of spontaneously generated coherence}\label{sec:sgc}

As mentioned above, our solution is valid throughout the full range
of magnetic field values, including the low magnetic field regime,
which was beyond the  scope of our earlier work.\cite{Barnes_PRL11}
This is achieved by taking fully into account the phenomenon of
spontaneously generated coherence (SGC), which is present in these
quantum dots and is most prominent at low magnetic fields. The
effect of SGC, first predicted theoretically in the early
1990s\cite{Javanainen_EL92} and about a decade later investigated
theoretically\cite{Economou_PRB05} and experimentally\cite{Dutt_PRL05} in
the context of optically controlled quantum dots, amounts to a
coherence term in the decay equations driven by the spontaneous
emission from the excited level. Though this may seem
counterintuitive at first, it is not difficult to understand if we
consider the limiting case of zero magnetic field in our system. In
that case, the only decay process is from state $|\bar T\rangle$ to
$|\bar x\rangle$ both in the lab and in the rotating frame.
Eqs.~(\ref{decaywsgc}) then reduce to a single nontrivial equation,
\beq \dot R_{\bar{x}\bar{x}} =\dot {\widetilde{R}}_{\bar{x}\bar{x}}
= 2\gamma R_{\bar{T}\bar{T}}. \label{eq:decay_Bzero} \eeq This
reflects the fact that in the $B=0$ limit, state $|x\rangle$ is
completely decoupled from the dynamics. When the field is switched
on, Eq.~(\ref{eq:decay_Bzero}) still holds in the lab frame, and the
additional terms in Eqs.~(\ref{decaywsgc}) arise from the
transformation to the rotating frame. We can understand intuitively
the origin of the SGC term from the fact that population terms in
one basis give rise to coherence terms in a different basis. Thus,
in the magnetic field basis $z$, the term $R_{\bar{x}\bar{x}}$ is a
linear combination of all four population and coherence terms
$R_{ij}$ with $i,j=|z\rangle, |\bar{z}\rangle$. Therefore, there is
a coherence term generated by spontaneous emission. This effect is
independent of basis, and in the $x$ basis and rotating frame, it
can be expressed as a coherence between states $|x\rangle$ and
$|\bar{x}\rangle$, as seen in Eqs.~(\ref{decaywsgc}).

To discuss the effects this term has on the nuclear dynamics, it is
useful to first think of the effect on the electron spin alone as
compared to the  absence of SGC. This was already discussed at
length in Ref.~\onlinecite{Economou_PRB05}, but it is worth summarizing
that discussion here for the sake of completeness. First, by
inspecting the decay equations and their solutions, we can
immediately see that SGC has the tendency to create electron
spin polarization along the $+y$ axis. This can also be seen through
a geometric picture of the spin as follows: the pulse removes part
of the spin vector pointing along $-x$ (how much depends on $q_o$).
The spin vector that remains along $x$ precesses about the $z$ axis
counterclockwise. For concreteness, consider a mixed initial state
and a pulse that is close to $\pi$, i.e., $q_o\sim 0$. There is then
a net spin vector component pointing along $+x$ which begins to
precess toward $+y$. As the spontaneous emission occurs, it can be
thought of as contributing small vectors that point toward $-x$
adding on to the `unexcited' part of the spin vector, which is now
in the $x,y>0$ quadrant. This process continues until the excited
state has fully decayed. The spontaneous emission in this case
partially opposes the generation of coherence by adding a coherent
component along $-x$. However, the spin component that is pointing
along $y$ is `protected', and thus in the rotating frame we can
think of a net $y$ component created by the total process of
excitation and spontaneous emission including SGC.

\begin{figure}[htp]
\;\;\;\quad\quad  \includegraphics[width=.8\columnwidth]{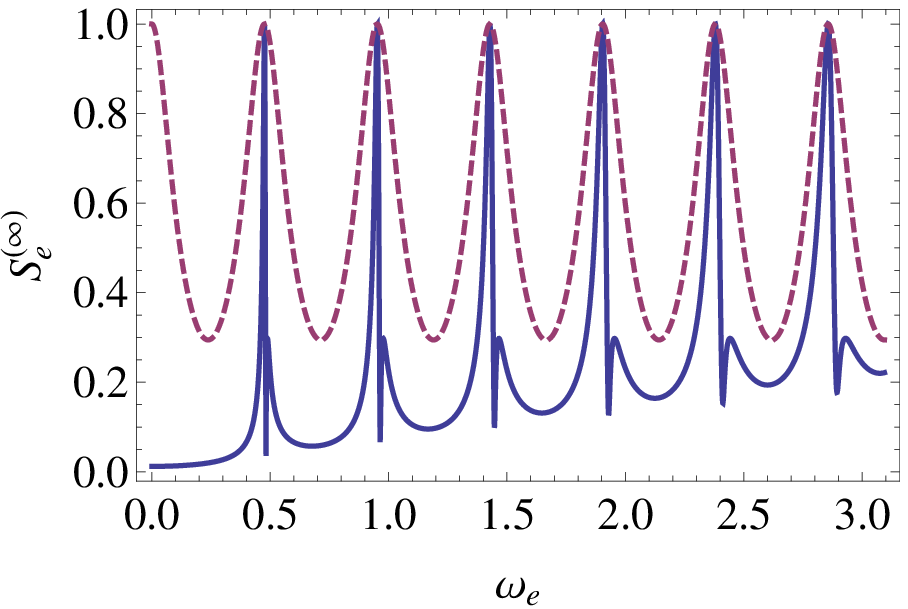}
  \includegraphics[width=0.84\columnwidth]{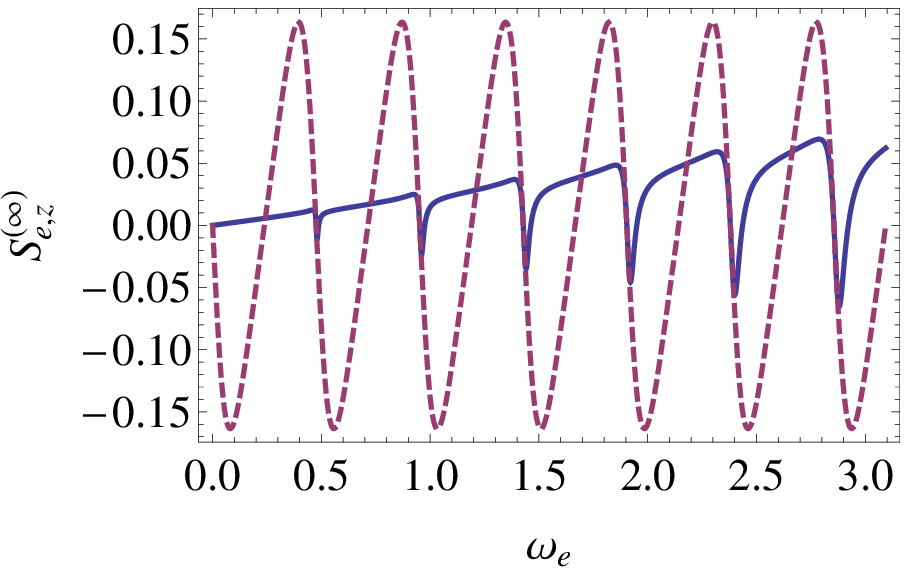}
  \caption{{\it Checking the effect of SGC}: Total electron steady state spin vector (top panel) and $z$ component (bottom panel) for $T_R=13.2$ns, $\phi=-\pi/2$, and $\gamma=0.5$GHz (solid), $\gamma=0$ (dashed). For $\gamma=0$, there is no SGC.}\label{fig:esstotnoSGC}
\end{figure}

\begin{figure*}
  \includegraphics[width=\textwidth]{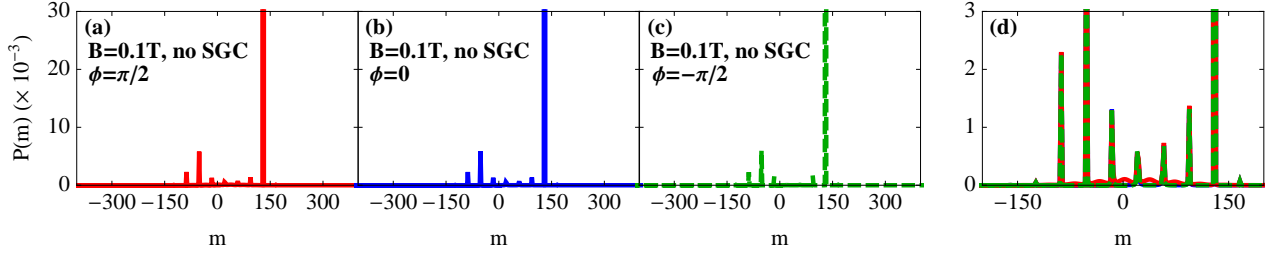}
  \caption{{\it Checking the effect of SGC}: (a-c) Nuclear spin polarization distribution for $T_R=13.2$ns, $NA=12.5$GHz, $N=3000$, $\gamma=0$, $q_o=0.3$, $\omega_n=0$, $B=0.1$T for various rotation angles. (d) Zoom-in of (a,b,c).}\label{fig:Pofm_noSGC}
\end{figure*}

\begin{figure}
  \includegraphics[width=0.8\columnwidth]{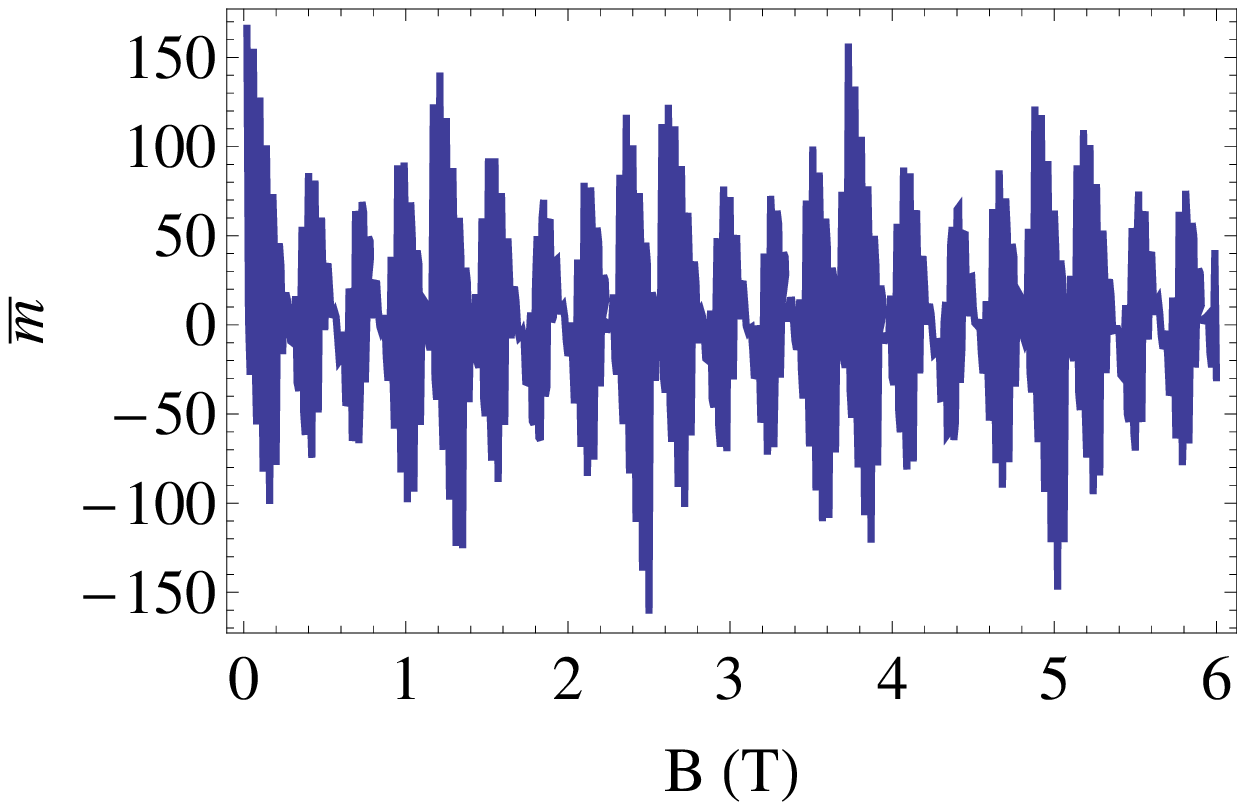}
  \caption{{\it Checking the effect of SGC}:  Average nuclear spin polarization versus magnetic field for $T_R=13.2$ns, $NA=12.5$GHz, $N=3000$, $\gamma=0$, $q_o=0.3$, $\omega_n=0$, and $\phi=\pi/2$.}\label{fig:mave_noSGC}
\end{figure}

The situation is more complicated when the pulse, in addition to
polarizing/depolarizing the spin vector, also rotates the spin.  This is
the case when $q_o\neq 0, \phi\neq 0$. The rotation is about the $x$
axis, so that the $y$ component of the spin vector due to SGC is
rotated to $z$. As a result, there is a persistent $z$ component of
the spin vector in the general case when both SGC and the
pulse-induced rotation are considered. This effect is evident in the
asymmetric form of the $z$ component shown in Fig. \ref{fig:esstotnoSGC}. The $z$ component of the
nuclear spin is itself a monotonically increasing function of the
electron spin $z$ component, Eq.~(\ref{snzwn0}). Therefore, SGC has
the effect of creating more nuclear polarization along $z$ on average, which in
turn controls the relative size of the nuclear spin flip rates in
the two directions (up to down vs down to up). We therefore expect
this to translate into a larger nuclear spin polarization relative
to the case of no SGC, which is what we see when we compare
Figs.~\ref{fig:Pofm}, \ref{fig:mave} and \ref{fig:Pofm_noSGC}, \ref{fig:mave_noSGC}.

\subsection{Spin echo}\label{sec:spinecho}

We now examine how the nuclear feedback mechanism is altered by the addition of an extra, unitary pulse in the middle of each period, i.e., at a time interval $T_R/2$ from the nonunitary pulse of the sequence. We specifically choose this additional pulse such that it implements a $\pi$ rotation of the electron spin around the $x$ axis. This can therefore be thought of as a spin echo sequence, and the pulse as an echo pulse. It is straightforward to include this pulse into our formalism by replacing the unitary part of the evolution $U_{hf}(T_R)$ in between pulses by
\begin{eqnarray}
U_{hf}(T_R/2) (R_x(\pi)\otimes \mathbbm{1}) U_{hf}(T_R/2),
\end{eqnarray}
where $R_x(\pi)=e^{-i \pi s_1}$ denotes the spin rotation
implemented by the echo pulse. In this case, the steady state of the
electron spin turns out to be \beq S_e^{SE}=(1,0,0). \eeq This steady
state coincides with that of the synchronized spins in the absence
of the echo pulse, so one may be tempted to think that the  dynamics
is trivial in the spin echo case and that no nuclear dynamics
occurs. This is in fact false; the nuclear dynamics and subsequent
feedback mechanism turn out to be distinct and interesting in the
presence of spin echo.

Following the same steps outlined in Appendix
\ref{app:nucspinSS_gen}, we find that the nuclear spin steady state
is trivial, $S_n^{(\infty)}=(0,0,0)$, while the nuclear spin  flip rates are
\beq \mathrm{w}_+ = \mathrm{w}_- = \frac{A^2 \sin^2(\omega_e
T_R/4)}{\omega_e^2 T_R}. \label{spinechorates} \eeq This result
clearly differs from what would be obtained from
Eqs.~(\ref{snzwn0})-(\ref{lambda2wn0}) and (\ref{singlenucleusw}) if we were to set $S_e=(1,0,0)$, with perhaps
the most striking difference being the extra factor of $1/2$ in the
argument of the sine. This indicates that the synchronized electron
Zeeman frequencies are no longer given by $2n\pi/T_R$ but instead by
$4n\pi/T_R$. The physical origin of this is that the evolution of
the electron-nuclear spin entanglement is modified by the echo
pulse, and in particular it is no longer the case that spins become
disentangled after a time span of $2\pi/\omega_e$; instead, this
disentanglement occurs after a time interval of $4\pi/\omega_e$.
Therefore, if the nonunitary pulse is applied at time
$t=2\pi/\omega_e$, the residual entanglement will polarize the
nuclear spin, whereas if it is applied at $t=4\pi/\omega_e$, no
polarization is produced.

\begin{figure}
  \includegraphics[width=\columnwidth]{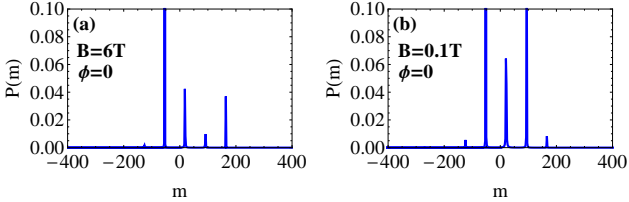}
  \caption{{\it Effect of spin echo pulse:} Nuclear spin polarization distribution for $T_R=13.2$ns, $NA=12.5$GHz, $N=3000$, $\gamma=0.5$GHz, $q_o=0.3$, $\omega_n=0$, $\phi=0$, and (a) $B=6$T, (b) $B=0.1$T.}\label{fig:Pofmse}
\end{figure}

\begin{figure}
  \includegraphics[width=0.8\columnwidth]{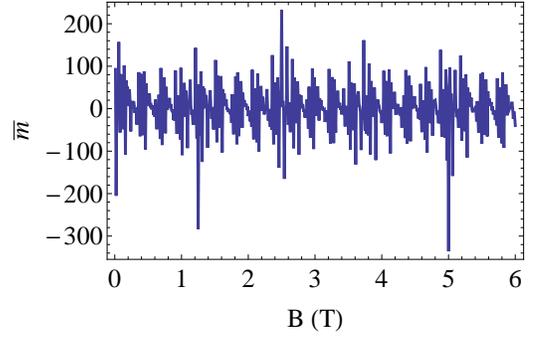}
  \caption{{\it Effect of spin echo pulse:} Average nuclear spin polarization distribution versus magnetic field for $T_R=13.2$ns, $NA=12.5$GHz, $N=3000$, $\gamma=0.5$GHz, $q_o=0.3$, $\omega_n=0$, $\phi=0$.}\label{fig:mavvsBse}
\end{figure}

Using the rates from Eq.~(\ref{spinechorates}), we calculate the
nuclear spin polarization distribution, and the results  are shown
in Fig.~\ref{fig:Pofmse}. As in the case without the echo pulse, we
find that the distribution exhibits a sequence of equally spaced
peaks, this time with a spacing period of $\Delta m=8/(A T_R)$,
twice as large as without echo. These peaks again indicate a
focusing effect, this time at the spin-echo synchronized Zeeman
frequencies, $4n\pi/T_R$. Since $\mathrm{w}_+ = \mathrm{w}_-$
regardless of the value of $\phi$ or any other parameters, the
physics of this focusing effect is analogous to the $\phi=0$ case in
the absence of spin echo. The equality of the flip rates also means
that the net nuclear spin polarization is minimal, as shown in
Fig.~\ref{fig:mavvsBse}.
The fact that $\mathrm{w}_+=\mathrm{w}_-$ allows for an explicit analytical solution of the polarization distribution
in the continuum limit, as shown in Appendix \ref{app:continuum}.

\section{Conclusions and outlook}

The field of nuclear spin control via the combination of coherent and incoherent driving of an electron spin confined in a quantum dot, while very vibrant and rapidly growing, is still at an early stage regarding a microscopic understanding of the key fundamental processes that occur in these systems. In this work, we presented a general formalism to treat the generation of DNP and its feedback effects on the electron spin both microscopically and self-consistently. Although these experiments are quite complex, we showed that by taking advantage of the separation of timescales that typically occur in these experimental setups and by employing powerful techniques from the field of quantum information, the theoretical description cannot only be rendered tractable but can yield analytical results that permit greater insight into the underlying physics. In particular, our formalism reveals the crucial role of electron-nuclear entanglement in the formation of DNP.

Our theory can be adapted in principle to any experimental setup where the electron is driven while interacting with the nuclear bath and with an additional reservoir. To demonstrate our theoretical framework, in this paper we analyzed in detail the mode locking experiments in which the electron spin is driven by a periodic train of fast circularly polarized laser pulses. We showed that our theory reproduces the main signatures of DNP in these experiments, namely the synchronization and anti-synchronization of the electron spin precession with the pulse repetition rate. Furthermore, our formalism predicts an enhancement of DNP at lower external magnetic fields due to the phenomenon of spontaneously generated coherence, an effect that was not included in previous treatments of the nuclear spin/DNP problem, and which has not yet been reported experimentally in this context to our knowledge. In addition, we applied our theory to the case where an extra spin echo pulse is inserted between every adjacent pair of mode locking pulses; this insertion constitutes the simplest implementation of dynamical decoupling in these experiments. Our results predict that the inclusion of the spin echo pulse both modifies the synchronization condition and significantly reduces the amount of DNP generated.

In this work we made certain assumptions and approximations. First and foremost, we exploited the separation of timescales in this problem, which enabled a Markovian approach. We argued that because processes leading to electron spin decoherence were ignored, the Markovian approximation here was actually \emph{more} appropriate compared to a non-Markovian treatment in which the full electron-nuclear spin correlations are retained. This is because the decoherence is fast compared to nuclear spin dynamics, implying that such correlations will decay before they become significant. Moreover, since the electron spin reaches its dynamical steady state quickly compared to the decoherence timescale, the electron will tend to remain in its steady state on average. These observations together suggest that the primary source of DNP feedback on the electron spin is through a modification of its precession frequency due to the Overhauser field of the nuclear spins; this is the type of feedback we have focused on in this work.

A complete treatment of the full dynamics of this problem would entail going beyond the Markovian limit by properly taking into account electron spin decoherence at a microscopic level. If this could be done, it would constitute an important breakthrough as it would lead to a formalism capable of describing any DNP experiment in quantum dots. However, this is a challenging problem as it would require abandoning the independent-nucleus approximation and including the full effect of the nuclear spin ensemble in the calculation of the electron spin steady state. Going beyond the independent-nucleus approximation would also allow us to investigate the role of inter-nuclear spin entanglement in the generation of DNP and in nuclear feedback effects. A promising approach to achieve this would be to incorporate techniques from the theory of generalized master equations\cite{Barnes_PRB11} into our formalism. These techniques are similar in spirit to the operator sum representation employed in this work in that they can offer a dramatic reduction in the effective size of the Hilbert space without invoking additional assumptions or approximations.  We leave the development and exploration of this more complete theoretical formalism to future work.

\bigskip
\centerline{\bf Acknowledgments}
\bigskip

We thank S. Carter and E.N. Economou for their careful reading of the manuscript and useful comments. This work was supported by LPS-CMTC (EB) and in part by ONR
(SEE).

\appendix

\section{Nuclear spin steady state and relaxation rate from perturbation theory} \label{app:nucspinSS_gen}

As discussed in Section~\ref{sec:nucspinSS_gen}, we can obtain analytical expressions for the nuclear spin steady state and relaxation rate by performing a perturbative expansion in the hyperfine flip-flop interaction (retaining the Overhauser part of the interaction to all orders). The first step is to expand $\mathcal{Y}_n$ in powers of the hyperfine flip-flop interaction:
\beq
\mathcal{Y}_n=\mathcal{Y}^{(0)}_{n}+\mathcal{Y}^{(1)}_{n}+\mathcal{Y}^{(2)}_{n}+...\label{expandYn}
\eeq
A similar expansion can be performed for the effective 4d spin vector of the nucleus:
\beq
\mathcal{S}_n=\mathcal{S}^{(0)}_{n}+\mathcal{S}^{(1)}_{n}+\mathcal{S}^{(2)}_{n}+...\label{expandSn}
\eeq
The goal of this appendix is to derive a formula for the nuclear spin relaxation rate and zeroth-order
steady state in terms of the first three terms in the expansion of $\mathcal{Y}_n$, Eq. (\ref{expandYn}). The key
observation that facilitates this derivation is that to zeroth order in the flip-flop term, the evolution of the nuclear spin is simple precession, at most modified by the effective magnetic field due to the electron spin component along the $z$ axis (the so-called Knight field). Thus, $\mathcal{Y}^{(0)}_{n}$ will have the general form
\begin{eqnarray}
\mathcal{Y}^{(0)}_n=\left[
\begin{array}{cccc}
1 & 0 & 0 & 0 \\
0 &  \mathcal{Y}^{(0)}_{n,xx} &  \mathcal{Y}^{(0)}_{n,xy} & 0 \\
0 &  \mathcal{Y}^{(0)}_{n,yx} &  \mathcal{Y}^{(0)}_{n,yy} & 0 \\
0 & 0 & 0 & 1
\end{array}
\right]. \label{yno}
\end{eqnarray}
Here, the first row is $(1,0,0,0)$ because this is generally the case for an evolution operator in the 4d SV representation.
The first column is $(1,0,0,0)$ because no polarization is generated---the nuclear spin evolution is unitary at zeroth order. The remaining
3$\times$3 submatrix implements (modified) precession in the $xy$ plane.
It will turn out that this generic form is already sufficiently restricted that we can make substantial progress
without specifying the explicit expressions for the $\mathcal{Y}^{(0)}_{n,xx}$, etc., or for
$\mathcal{Y}^{(1)}_{n}$ and $\mathcal{Y}^{(2)}_{n}$.

The nuclear spin steady state is defined as the solution to the following eigenvalue equation:
\beq
(\mathbbm{1}-\mathcal{Y}_n)\mathcal{S}_n=\lambda \mathcal{S}_n,
\eeq
with $\lambda=0$. We have kept $\lambda$ in this equation since we will need to consider nonzero eigenvalues as well in order to obtain the relaxation time. Using the above expansions and equating terms occurring at the same level, we find
\bea
(\mathbbm{1}-\mathcal{Y}^{(0)}_{n})\mathcal{S}^{(0)}_{n}&=&0,\nn\\
(\mathbbm{1}-\mathcal{Y}^{(0)}_{n})\mathcal{S}^{(1)}_{n}&=&(\mathcal{Y}^{(1)}_{n}+\lambda_1) \mathcal{S}^{(0)}_{n},\label{perturbativeeqns}\\
(\mathbbm{1}-\mathcal{Y}^{(0)}_{n})\mathcal{S}^{(2)}_{n}&=&(\mathcal{Y}^{(2)}_{n}+\lambda_2) \mathcal{S}^{(0)}_{n}+(\mathcal{Y}^{(1)}_{n}+\lambda_1) \mathcal{S}^{(1)}_{n}\nn.
\eea
In the first of these equations, we have taken the liberty of setting $\lambda_0=0$ since the relaxation time will be related to the smallest eigenvalue of $\mathbbm{1}-\mathcal{Y}_n$. More specifically, it is apparent from Eq. (\ref{yno}) that at zeroth order, two of the eigenvalues of $\mathbbm{1}-\mathcal{Y}^{(0)}_{n}$ vanish, so that we have a degenerate perturbation theory. One of these eigenvalues will remain zero at all orders of the perturbative expansion, and the corresponding eigenvector is the effective steady state of the nucleus. The second zero eigenvalue will receive corrections at higher orders. Since these corrections will be proportional to the hyperfine coupling, this will then correspond to the smallest nonzero eigenvalue of $\mathbbm{1}-\mathcal{Y}_n$ when the coupling is sufficiently small.

The first equation in (\ref{perturbativeeqns}) states that the zeroth-order eigenvectors $\mathcal{S}^{(0)}_{n}$ with vanishing eigenvalues live in the null space of $\mathbbm{1}-\mathcal{Y}^{(0)}_{n}$. It is clear from Eq. (\ref{yno}) that the null space of $\mathbbm{1}-\mathcal{Y}^{(0)}_n$ is spanned by the vectors $v_0\equiv(1,0,0,0)$ and $v_1\equiv(0,0,0,1)$. Since the first component of the steady state $\mathcal{S}_{n,ss}$ must be fixed at 1 (see the discussion following Eq. (\ref{espin})), we may write for the zeroth-order steady state
\beq
\mathcal{S}^{(0)}_{n,ss}=(1,0,0,\xi),
\eeq
where $\xi$ is a constant. The fact that the value of $\xi$ is not determined by the zeroth-order equation means that in the absence of hyperfine flip-flops, the nuclear spin steady state is not unique and depends on the initial state. When hyperfine flip-flops are included by taking into account the higher-order equations in (\ref{perturbativeeqns}), the value of $\xi$ becomes fixed, and the steady state is unique. To see this, we need to solve both the first and second-order equations in (\ref{perturbativeeqns}).

Consider the first-order equation in (\ref{perturbativeeqns}). We can dot both sides of this equation by the vectors $v_0$ and $v_1$ to obtain
\beq
v_0(\mathcal{Y}^{(1)}_{n}+\lambda_1)\mathcal{S}^{(0)}_{n}=0,\quad v_1(\mathcal{Y}^{(1)}_{n}+\lambda_1) \mathcal{S}^{(0)}_{n}=0.
\eeq
It is generally the case that the components $\mathcal{Y}^{(1)}_{n,00}$ and $\mathcal{Y}^{(1)}_{n,zz}$ are zero, so that $v_{0}\mathcal{Y}^{(1)}_{n}v_{0}=v_{1}\mathcal{Y}^{(1)}_{n}v_{1}=0$, implying $\lambda_1=0$. The reason $\mathcal{Y}^{(1)}_{n,00}$ vanishes is due to the requirement that $\mathcal{Y}_{n,00}=1$, which must hold for all evolution operators in the 4d SV representation, and which is already satisfied by $\mathcal{Y}^{(0)}_{n,00}$. The component $\mathcal{Y}^{(1)}_{n,zz}$ vanishes because populations are unaltered in first-order perturbation theory. Since $\lambda_1=0$, the relaxation rate will be at least second-order in the hyperfine coupling. It is not difficult to directly solve the first-order equation in (\ref{perturbativeeqns}), with the result
\beq
\mathcal{S}^{(1)}_{n}=p_1+bv_1,\label{Sn1sol}
\eeq
where $p_1$ is a vector of the form $p_1=(0,p_{12},p_{13},0)$ which can be obtained explicitly from the following formula:
\beq
p_1=\left[\begin{matrix} 1&0&0&0 \cr 0& X_{xx} & X_{xy} &0 \cr 0& X_{yx} & X_{yy} &0 \cr 0&0&0&1 \end{matrix}\right]\mathcal{Y}^{(1)}_n(1,0,0,\xi),
\eeq
where the 2$\times$2 matrix $X$ is defined as
\beq
X\equiv\left[\begin{matrix} 1-\mathcal{Y}^{(0)}_{n,xx} & \mathcal{Y}^{(0)}_{n,xy} \cr \mathcal{Y}^{(0)}_{n,yx} & 1-\mathcal{Y}^{(0)}_{n,yy}\end{matrix}\right]^{-1}.
\eeq
It should be noted that $p_1$ is a function of the constant $\xi$. The additional constant $b$ appearing in Eq.~(\ref{Sn1sol}) is arbitrary. We do not include a term proportional to $v_0$ as well because this would violate the constraint that the first component of the 4d spin vector is fixed to 1 since we have already set the first component of the zeroth-order steady state, $\mathcal{S}^{(0)}_{n,ss}$, to 1.

Dotting both sides of the second-order equation in (\ref{perturbativeeqns}) by $v_1$ gives
\beq
v_1\mathcal{Y}^{(1)}_{n}p_1+v_1(\mathcal{Y}^{(2)}_{n}+\lambda_2)\mathcal{S}^{(0)}_{n}=0.\label{2ndordereqn}
\eeq
This equation has two solutions:
\beq
\mathcal{S}^{(0)}_{n}=(1,0,0,\xi^*),\qquad \lambda_2=0,\label{nucss}
\eeq
and
\beq
\mathcal{S}^{(0)}_n=(0,0,0,1),\qquad \lambda_2=\lambda_2^*.\label{eigrelax}
\eeq
The first solution, Eq.~(\ref{nucss}), is the nuclear steady state spin vector, while the second solution, Eq.~(\ref{eigrelax}), gives the nuclear spin relaxation rate (the rate at which the nuclear spin reaches its steady state):
\beq
\gamma_n=\lambda_2^*/T_R.\label{nucrelax}
\eeq
As we anticipated, the nuclear spin steady state is unique in the presence of hyperfine flip-flops. The explicit expressions for $\xi^*, \lambda_2^*$ depend on the particular control sequence.

\begin{widetext}
\section{Effective nuclear spin evolution to second order in hyperfine flip-flops for single-pulse-per-period driving}\label{app:Yn}
As explained in the previous appendix, the effective nuclear spin evolution operator in the spin vector representation can be expanded to second order in the hyperfine flip-flop interaction:
\beq
\mathcal{Y}_n=\mathcal{Y}^{(0)}_{n}+\mathcal{Y}^{(1)}_{n}+\mathcal{Y}^{(2)}_{n}+...
\eeq
In this expansion, we are working in the Markovian limit, and we are retaining the Overhauser part of the interaction to all orders. The explicit form of the zeroth-order evolution in the case of a single pulse per driving period is
\beq
\mathcal{Y}^{(0)}_{n}=\left[
\begin{array}{cccc}
 1 & 0 & 0 & 0 \\
 0 & \cos \left(\frac{A T_R}{2}\right) \cos \left(T_R \omega _n\right)-\sin \left(\frac{A T_R}{2}\right) \sin \left(T_R \omega _n\right) S_{e,z}
   & -\cos \left(\frac{A T_R}{2}\right) \sin \left(T_R \omega _n\right)-\cos \left(T_R \omega _n\right) \sin \left(\frac{A T_R}{2}\right)
   S_{e,z} & 0 \\
 0 & \cos \left(\frac{A T_R}{2}\right) \sin \left(T_R \omega _n\right)+\cos \left(T_R \omega _n\right) \sin \left(\frac{A T_R}{2}\right) S_{e,z}
   & \cos \left(\frac{A T_R}{2}\right) \cos \left(T_R \omega _n\right)-\sin \left(\frac{A T_R}{2}\right) \sin \left(T_R \omega _n\right) S_{e,z}
   & 0 \\
 0 & 0 & 0 & 1 \\
\end{array}
\right],
\eeq
and the nonzero components of the first and second-order contributions are
\bea
\mathcal{Y}^{(1)}_{n,x0}&=&\frac{A \sin \left(\frac{A T_R}{2}\right) \sin \left(\frac{1}{2} T_R \left(\omega _e-\omega _n\right)\right) \left[S_{e,x} \cos
   \left(\frac{1}{2} T_R \left(\omega _e+\omega _n\right)\right)-S_{e,y} \sin \left(\frac{1}{2} T_R \left(\omega _e+\omega
   _n\right)\right)\right]}{\omega _e-\omega _n},\nn\\
\mathcal{Y}^{(1)}_{n,xz}&=&\frac{A \cos \left(\frac{A T_R}{2}\right) \sin \left(\frac{1}{2} T_R \left(\omega _e-\omega _n\right)\right) \left[S_{e,x} \sin
   \left(\frac{1}{2} T_R \left(\omega _e+\omega _n\right)\right)+S_{e,y} \cos \left(\frac{1}{2} T_R \left(\omega _e+\omega
   _n\right)\right)\right]}{\omega _e-\omega _n},\nn\\
\mathcal{Y}^{(1)}_{n,y0}&=&\frac{A \sin \left(\frac{A T_R}{2}\right) \sin \left(\frac{1}{2} T_R \left(\omega _e-\omega _n\right)\right) \left[S_{e,x} \sin
   \left(\frac{1}{2} T_R \left(\omega _e+\omega _n\right)\right)+S_{e,y} \cos \left(\frac{1}{2} T_R \left(\omega _e+\omega
   _n\right)\right)\right]}{\omega _e-\omega _n},\nn\\
\mathcal{Y}^{(1)}_{n,yz}&=&-\frac{A \left(1+e^{i A T_R}\right) e^{-\frac{1}{2} i T_R \left(A+\omega _e+\omega _n\right)} \sin \left(\frac{1}{2} T_R \left(\omega _e-\omega
   _n\right)\right) \left[\left(S_{e,x}+i S_{e,y}\right) e^{i T_R \left(\omega _e+\omega _n\right)}+S_{e,x}-i S_{e,y}\right]}{4 \left(\omega
   _e-\omega _n\right)},\nn\\
\mathcal{Y}^{(1)}_{n,zx}&=&-\frac{A \left[-S_{e,x}\cos \left(T_R \left(\omega _e-\omega _n\right)\right)+S_{e,y} \sin \left(T_R \left(\omega _e-\omega
   _n\right)\right)+S_{e,x}\right]}{2 \left(\omega _e-\omega _n\right)},\nn\\
\mathcal{Y}^{(1)}_{n,zy}&=&\frac{A \left\{S_{e,x} \sin \left(T_R \left(\omega _e-\omega _n\right)\right)+S_{e,y} \left[\cos \left(T_R \left(\omega _e-\omega
   _n\right)\right)-1\right]\right\}}{2 \left(\omega _e-\omega _n\right)},
\eea
\bea
\mathcal{Y}^{(2)}_{n,xx}&=&\frac{A^2}{4 \left(\omega
   _e{-}\omega _n\right){}^2}\bigg\{S_{e,z} \sin \left(\frac{A T_R}{2}\right) \left[T_R \left(\omega _e-\omega _n\right) \cos \left(\omega _n T_R\right)-\sin
   \left(\omega _e T_R\right)+\sin \left(\omega _n T_R\right)\right]\nn\\&&+\cos \left(\frac{A T_R}{2}\right) \left[T_R \left(\omega _e-\omega
   _n\right) \sin \left(\omega _n T_R\right)+\cos \left(\omega _e T_R\right)-\cos \left(\omega _n T_R\right)\right]\bigg\},\nn\\
\mathcal{Y}^{(2)}_{n,xy}&=&\frac{A^2}{4 \left(\omega
   _e-\omega _n\right){}^2} \bigg\{S_{e,z} \sin \left(\frac{A T_R}{2}\right) \left[T_R \left(\omega _n-\omega _e\right) \sin \left(\omega _n T_R\right)-\cos
   \left(\omega _e T_R\right)+\cos \left(\omega _n T_R\right)\right]\nn\\&&+\cos \left(\frac{A T_R}{2}\right) \left(T_R \left(\omega _e-\omega
   _n\right) \cos \left(\omega _n T_R\right)-\sin \left(\omega _e T_R\right)+\sin \left(\omega _n T_R\right)\right]\bigg\},\nn\\
\mathcal{Y}^{(2)}_{n,yx}&=&\frac{A^2}{4 \left(\omega
   _e-\omega _n\right){}^2} \bigg\{S_{e,z} \sin \left(\frac{A T_R}{2}\right) \left[T_R \left(\omega _e-\omega _n\right) \sin \left(\omega _n T_R\right)+\cos
   \left(\omega _e T_R\right)-\cos \left(\omega _n T_R\right)\right]\nn\\&&+\cos \left(\frac{A T_R}{2}\right) \left[T_R \left(\omega _n-\omega
   _e\right) \cos \left(\omega _n T_R\right)+\sin \left(\omega _e T_R\right)-\sin \left(\omega _n T_R\right)\right]\bigg\},\nn\\
\mathcal{Y}^{(2)}_{n,yy}&=&\frac{A^2}{4 \left(\omega
   _e-\omega _n\right){}^2} \bigg\{S_{e,z} \sin \left(\frac{A T_R}{2}\right) \left[T_R \left(\omega _e-\omega _n\right) \cos \left(\omega _n T_R\right)-\sin
   \left(\omega _e T_R\right)+\sin \left(\omega _n T_R\right)\right]\nn\\&&+\cos \left(\frac{A T_R}{2}\right) \left[T_R \left(\omega _e-\omega
   _n\right) \sin \left(\omega _n T_R\right)+\cos \left(\omega _e T_R\right)-\cos \left(\omega _n T_R\right)\right]\bigg\},\nn\\
\mathcal{Y}^{(2)}_{n,z0}&=&\frac{A^2 S_{e,z} \sin ^2\left(\frac{1}{2} T_R \left(\omega _e-\omega _n\right)\right)}{\left(\omega _e-\omega _n\right){}^2},\nn\\
\mathcal{Y}^{(2)}_{n,zz}&=&\frac{A^2 \left(\cos \left(T_R \left(\omega _e-\omega _n\right)\right)-1\right)}{2 \left(\omega _e-\omega _n\right){}^2}.
\eea
\end{widetext}

\section{Nuclear spin steady state and relaxation rate for single-pulse-per-period driving}\label{app:nssrr}

In the case of driving with a single pulse per period as considered in Section \ref{sec:drivingwsimpleperiodicpulse_gen} and which is relevant for the mode locking experiments analyzed in detail in Section \ref{ML}, the explicit forms of the nuclear spin steady and relaxation rate for arbitrary nuclear Zeeman energy $\omega_n$ are
\bea
\mathcal{S}^{(\infty)}_{n}&=&(1,0,0,\xi^*),\nn\\
\gamma_n&=&\lambda_2^*/T_R,
\eea
with
\bea
\xi^*&=&S_{n,z}^{(\infty)}=\frac{C}{D},\nn\\
\lambda_2^*&=&\frac{A^2}{4 \left(\omega _e-\omega _n\right){}^2} \left(2-F/G\right),
\eea
where
\begin{widetext}
\bea
C&=&-2 e^{i T_R \left(A+\omega _n\right)} \bigg\{2 \left(S_{e,z}^2+S_e^2\right) \sin \left(\frac{A T_R}{2}\right) \sin \left(\omega _n
   T_R\right)\nn\\&&+S_{e,z} \left[-\left(S_e^2-1\right) \cos \left(A T_R\right)-4 \cos \left(\frac{A T_R}{2}\right) \cos \left(\omega _n
   T_R\right)+S_e^2+3\right]\bigg\},
\eea
\bea
D&=&\left[\left(S_{e,z}-2\right) S_{e,z}-S_e^2+2\right] e^{\frac{1}{2} i T_R \left(A+4 \omega _n\right)}+2
   \left(-2 S_{e,z}^2+S_e^2-3\right) e^{i T_R \left(A+\omega _n\right)}\nn\\&&+\left[S_{e,z} \left(S_{e,z}+2\right)-S_e^2+2\right] e^{\frac{1}{2} i T_R \left(3 A+4 \omega _n\right)}+e^{\frac{3}{2} i A T_R} \left[\left(S_{e,z}-2\right) S_{e,z}-S_e^2+2\right]\nn\\&&+e^{\frac{1}{2} i A T_R}
   \left[S_{e,z} \left(S_{e,z}+2\right)-S_e^2+2\right]+\left(S_e^2-1\right) e^{i T_R \left(2 A+\omega _n\right)}+\left(S_e^2-1\right) e^{i
   \omega _n T_R},
\eea
\bea
F&=&2 (\xi-1)S_{e,\perp}^2\cos \left(\frac{1}{2} T_R \left(A-2 \omega _n\right)\right)+2 (\xi+1)
   S_{e,\perp}^2 \cos \left(\frac{1}{2} T_R \left(A+2 \omega _n\right)\right)\nn\\&&+\left[-(\xi-1) S_{e,\perp}^2-2
   S_{e,z}+2\right] \cos \left(\frac{1}{2} T_R \left(A+2 \omega _e-4 \omega _n\right)\right)+\left[(\xi-S_{e,z}) S_{e,\perp}^2
   +S_{e,z}^2-1\right] \cos \left(T_R \left(A+\omega _e-\omega _n\right)\right)\nn\\&&+\left[(\xi
   -S_{e,z}) S_{e,\perp}^2+S_{e,z}^2-1\right] \cos \left(T_R \left(A-\omega _e+\omega
   _n\right)\right)-\left[(\xi+1) S_{e,\perp}^2-2 \left(S_{e,z}+1\right)\right] \cos \left(\frac{1}{2} T_R \left(A-2 \omega _e+4
   \omega _n\right)\right)\nn\\&&-2 \cos \left(\frac{A T_R}{2}\right) \cos \left(\omega _e T_R\right) \left[\xi
   S_{e,\perp}^2-2\right]-2 S_{e,\perp}^2 \left[\cos \left(A T_R\right)
   \left(\xi-S_{e,z}\right)+S_{e,z}+\xi\right]\nn\\&&+2 \cos \left(T_R \left(\omega _e-\omega _n\right)\right) \left[\xi
   S_{e,\perp}^2+S_{e,z} \left(S_{e,\perp}^2-S_{e,z}\right)-3\right]+2 \sin \left(\frac{A T_R}{2}\right) \sin
   \left(\omega _e T_R\right) \left(S_{e,\perp}^2-2 S_{e,z}\right),
\eea
\bea
G&=&-4 S_{e,z} \sin \left(\frac{A T_R}{2}\right) \sin \left(\omega _n
   T_R\right)+\left(S_{e,z}^2-1\right) \cos \left(A T_R\right)-S_{e,z}^2+4 \cos \left(\frac{A T_R}{2}\right) \cos \left(\omega _n
   T_R\right)-3.
\eea
\end{widetext}
In the above expressions, we have compressed the notation for the electron steady state $S_{e,i}^{(\infty)}\to S_{e,i}$ for the sake of brevity, and we have defined $S_e^2\equiv S_{e,x}^2+S_{e,y}^2+S_{e,z}^2$.

\section{Derivation of flip rate expression}\label{app:fliprates}

At leading order in the hyperfine coupling, the nuclei are
essentially independent of each other, and we may estimate the flip
rates by using the solution we have  obtained for the single nucleus
problem. For a single nucleus, we may write \beq {dP_\uparrow\over
dt}=-\mathrm{w}_-P_\uparrow+\mathrm{w}_+P_\downarrow, \eeq where
$P_\uparrow$ is the probability that the nucleus is aligned with the
magnetic field and $P_\downarrow=1-P_\uparrow$ is the probability
that it lies antiparallel to the magnetic field. In terms of the
nuclear spin vector component along the magnetic field direction,
$S_{n,z}$, these probabilities are given by \beq
P_\uparrow={1\over2}(1+S_{n,z}),\qquad
P_\downarrow={1\over2}(1-S_{n,z}). \eeq Therefore, we have \beq
{d\over
dt}S_{n,z}=-(\mathrm{w}_++\mathrm{w}_-)S_{n,z}+\mathrm{w}_+-\mathrm{w}_-.
\eeq The solution to this equation is easily obtained: \beq
S_{n,z}(t) = \left[S_{n,z}(0)-{\mathrm{w}_+-\mathrm{w}_-\over
\mathrm{w}_++\mathrm{w}_-}\right]e^{-(\mathrm{w}_++\mathrm{w}_-)t}+{\mathrm{w}_+-\mathrm{w}_-\over
\mathrm{w}_++\mathrm{w}_-}.\label{rateEqnForOneNucleus} \eeq We may
then compute the flip rates $\mathrm{w}_\pm$ by comparing this with
our coarse-grained solution from Eq.~(\ref{coarseSn}): \beq
\mathcal{S}_n(t)=e^{(\mathcal{Y}_n-1)
t/T_R}\mathcal{S}_n(0).\label{coarseSnii} \eeq To facilitate the
comparison, we expand the initial state as a linear combination of
the eigenvectors of $1-\mathcal{Y}_n$: \beq
\mathcal{S}_n(0)=\mathcal{S}_n^{(\infty)}+\sum_{i=1}^3c_i\mathcal{V}_i.\label{initexpand}
\eeq Here, we have set the coefficient of the steady state
$\mathcal{S}_n^{(\infty)}$ to 1 since the first component of
$\mathcal{S}_n(0)$ must be 1, and the first component of
$\mathcal{S}_n^{(\infty)}$ is already 1. (Consequently, it must be
the case that the first components of each of the $\mathcal{V}_i$
are all zero.) Plugging Eq.~(\ref{initexpand}) into
Eq.~(\ref{coarseSnii}), we find \beq
\mathcal{S}_n(t)=\mathcal{S}_n^{(\infty)}+\sum_{i=1}^3c_ie^{-\mu_it/T_R}\mathcal{V}_i,\label{Sntexpand}
\eeq where the $\mu_i$ are the eigenvalues of $1-\mathcal{Y}_n$
corresponding to the $\mathcal{V}_i$.

At this point, we use the fact that the null space of
$1-\mathcal{Y}_n^{(0)}$ is two-fold degenerate and spanned by the
vectors $v_0=(1,0,0,0)$ and $v_1=(0,0,0,1)$, as discussed in Appendix \ref{app:nucspinSS_gen}. Two of
the four eigenvectors will therefore have a vanishing eigenvalue at
zeroth order in perturbation theory. One of these eigenvectors is
the steady state $\mathcal{S}_n^{(\infty)}$, and we choose the other
to be $\mathcal{V}_3$. This immediately implies that the steady
state has the form $\mathcal{S}_n^{(\infty)}=(1,0,0,\xi^*)$, while
$\mathcal{V}_3=(0,0,0,1)$ at zeroth order since the first component
of $\mathcal{S}_n^{(\infty)}$ is 1, while that of $\mathcal{V}_3$
has to be zero. Since the null space is orthogonal to the row space
spanned by $\mathcal{V}_1$ and $\mathcal{V}_2$, these vectors must
be orthogonal to $v_0$ and $v_1$ at zeroth order and therefore do
not have $z$-components at zeroth order. Taking the $z$-component of
Eq.~(\ref{Sntexpand}), we find that the zeroth-order nuclear spin
$z$-component is given by \beq
\mathcal{S}_{n,z}(t)=S_{n,z}^{(\infty)}+c_3e^{-\mu_3t/T_R}. \eeq
Identifying $\mu_3/T_R$ as the nuclear spin relaxation rate
$\gamma_n$, and rewriting $c_3$ in terms of the initial value
$S_{n,z}(0)$, we have \beq S_{n,z}(t) =
\left[S_{n,z}(0)-S_{n,z}^{(\infty)}\right]e^{-\gamma_nt}+S_{n,z}^{(\infty)}.\label{cgsoln}
\eeq Comparing this equation with Eq.~(\ref{rateEqnForOneNucleus})
gives \beq \mathrm{w}_\pm=\gamma_n(1\pm S_{n,z}^{(\infty)})/2, \eeq
which is quoted in Eq.~(\ref{singlenucleusw}).

\section{Continuum of rate equation}\label{app:continuum}

Starting from the recursion formula for the nuclear spin polarization distribution,
\begin{eqnarray}
P(m)= \frac{N-m+2}{N+m}~\frac{w_+(m-2)}{w_-(m)} P(m-2),
\end{eqnarray}
we can take the continuum limit by rewriting this as
\bea
&&w_-(m+2)P(m+2)-w_-(m)P(m)\nn\\&&=\frac{N-m}{N+m+2}w_+(m)P(m)-w_-(m)P(m).
\eea
Defining the function $\Phi(m)\equiv w_-(m)P(m)$, we can interpret the left-hand side as the derivative of $\Phi$
in the continuum limit:
\beq
\Phi'(m)=\frac{1}{2}\left[\frac{N-m}{N+m+2}\frac{w_+(m)}{w_-(m)}-1\right]\Phi(m).\label{Phieqn}
\eeq
This equation is easily integrated, with the result
\beq
\Phi(m)=C\exp\left(\frac{1}{2}\int_{-N}^m dm'\left[\frac{N-m'}{N+m'+2}\frac{w_+(m')}{w_-(m')}-1\right]\right).
\eeq
The constant $C$ is determined by the normalization of $P(m)$:
\beq
C=\left[\sum_m \frac{\exp\left(\frac{1}{2}\int_{-N}^m dm'\left[\frac{N-m'}{N+m'+2}\frac{w_+(m')}{w_-(m')}-1\right]\right)}{w_-(m)}\right]^{-1}.
\eeq
In the special case where the flip rates are equal, $w_+(m)=w_-(m)$, the polarization distribution reduces to
\beq
P(m)=\frac{2^{-1-m-N}Ce^{-m-N}(2+m+N)^{1+N}}{w_-(m)}.
\eeq
It is clear from this expression that the zeros of $w_-(m)$ give rise to peaks in $P(m)$. For the original mode locking experiment, $w_-(m)\sim\sin^2((\omega_e+Am/2)T_R/2)$, while in the case of spin echo, $w_-(m)\sim\sin^2((\omega_e+Am/2)T_R/4)$, immediately implying that $P(m)$ will exhibit a comb-like structure in both cases.


\begin{thebibliography}{49}
\bibitem{foot1}  An alternative approach to the mode locking experiment was recently given in Ref. \onlinecite{Glazov_PRB12}.
\bibitem{foot2} A similar model was recently considered in Ref. \onlinecite{Wu_arxiv13}.
\end{thebibliography}

\end{document}